\documentclass{appolb}
\usepackage{graphicx}
\usepackage{slashed}

\def\be{\begin{equation}}
\def\ee{\end{equation}}
\def\ba{\begin{eqnarray}}
\def\ea{\end{eqnarray}}

\def\sfrac#1#2{{\textstyle \frac{#1}{#2}}}

\newcommand{\ms}{\scriptscriptstyle}

\usepackage{graphics}
\usepackage{epsf}
\usepackage{amsmath}
\usepackage{amssymb}

\usepackage{mathabx}
\usepackage{bm}

\usepackage{hyperref}
\usepackage{url}

\usepackage{xcolor}
\newcommand{\B}{\color{blue}}

\begin{document}
\title{Nucleon to Roper transition amplitudes and electromagnetic form factors%
  %\thanks{Fest Roper volume at Acta Physica Polonica B, edited by Michal Praszalowicz and Igor Strakovsky}%
  \thanks{Dedicated to the 90th birthday of L.~D.~Roper.}
}
\author{G.~Ramalho, 
\address{Department of Physics and OMEG Institute, 
Soongsil University, \\
Seoul 06978, Republic of Korea}
\\[3mm]
}
\maketitle
\begin{abstract}
  The second excitation of the nucleon, the Roper, has properties differentiated from other low-lying nucleon resonances.
  Their properties challenge our understanding of the structure of the baryons in terms of the degrees of freedom from QCD.
  In the present work we discuss the properties of the Roper resonance and the nucleon to Roper electromagnetic transition, based on the quark degrees of freedom, that are expected to dominate for large square momentum transfer $Q^2$.
 We also discuss the analytic structure of the transition amplitudes in the low-$Q^2$ region, and how the contributions of baryon-meson states can help to describe the low and intermediate $Q^2$ data, and the nature of the Roper.
\end{abstract}

\section{Introduction \label{secIntro}}

David Roper contributed significantly to the study of nucleon resonances and to the discovery of the second excitation of the nucleon.
In the spectrum of the nucleon excitations ($N^\ast$), reported by Particle Data Group~\cite{PDG2024}, the first nucleon resonance is the $\Delta(1232)\frac{3}{2}^+$, and the Roper, now labeled as the $N(1440)\frac{1}{2}^+$ state, appears as the next excited state.
Contrary to the $\Delta(1232)$, identified clearly as an isolated peak on the $\gamma^\ast N \to \pi N$ cross sections, the $N(1440)$ state is more deceptive.
The $N(1440)$ state was found by David Roper and collaborators in the phase shift analysis of the elastic $\pi N$ scattering~\cite{Roper64a,Roper65a}.
Traces of the $N(1440)$ state can be better found in the two pion production channels
($N^\ast \to \pi \pi N$)~\cite{PPNP2024,NSTAR,Aznauryan12a,CLAS09,CLAS12,CLAS16a,Mokeev23a}.

In the $\gamma^\ast N \to \pi N$ cross sections, $N(1440)$ appears at the threshold of the second resonance region, a region dominated by the $N(1520)\frac{3}{2}^-$ and $N(1535)\frac{1}{2}^-$ resonances.
The $N(1440)$ contribution for the $\pi N$ channel is, however, overshadowed in the presence of those other resonances.
Evidence of the effect of the Roper can also be observed on $N N \to NN \pi \pi$ reactions~\cite{Hernandez02a,CELSIUS08a,ARuso98a,Cao10a}, and on the $N^\ast$ contributions to the inclusive distribution functions of the nucleon~\cite{Blin19a}.
In contrast, the impact of the Roper on the nucleon Compton scattering is small~\cite{Eichmann18}.

As discussed plentifully in the literature of the last 50 years, the nature of the Roper resonance has been a long-standing mystery.
Within the quark model framework, the resonance quantum numbers are consistent with the first radial excitation of the nucleon $N(940)\frac{1}{2}^+$.
The measured mass and decay width are, however, difficult to explain in terms of the valence quark degrees of freedom~\cite{Crede13a,Capstick86b,Isgur79b,Burkert19a}.
Models based exclusively on valence quarks predict a large mass, while the decay rates have significant contributions from $\pi \pi N$ states, including $\pi \Delta$ and $\sigma N$ channels~\cite{PDG2024}.
Studies of the nature of the Roper assume different compositions for the resonance, including gluon excitations~\cite{Li92a},
strong $\sigma N$ contributions~\cite{Alberto01a}, the interpretation
of the system as a breathing mode of the nucleon (quark-soliton model)~\cite{Kaulfuss85a,Mattis85a}, 
or that it is a dynamically generated resonance~\cite{Krehl00a,Ronchen13a,Doring09a,Wang24,Doring25a,Golli18a}.
Discussions on the subject can be found in other articles of David Roper's special volume.
For more details, we recommend the reviews on Refs.~\cite{PPNP2024,Aznauryan12a,Burkert19a}.

Our interest for the Roper was triggered by the experimental developments on the studies of the $\gamma^\ast N \to N^\ast$ transitions at Jefferson Lab (JLab-6 GeV).
More specifically by the measurements of the $\gamma^\ast N \to N(1440)$ helicity amplitudes above  $Q^2=1$ GeV$^2$, up to $Q^2= 4.3$ GeV$^2$~\cite{CLAS09,CLAS08}.
The new measurements re-initiated the discussion about the dynamical structure of the $N(1440)$ state.

Our approach to the Roper was motivated by the success of our previous studies, using a quark model framework, on the nucleon elastic form factors~\cite{Nucleon}, and on the $\gamma^\ast N \to \Delta(1232)$ transition form factors~\cite{NDelta,NDeltaD,Lattice,LatticeD,Timelike,Timelike2,Siegert1,Siegert3,Delta-shape}.
In those studies we considered the covariant spectator quark model, where the structure of the wave functions of the baryons is established by the SU(6) quark-flavor symmetry, and the shape of the radial wave functions is determined by symmetry and by phenomenology~\cite{Nucleon,Omega,NSTAR2017,Nucleon2}.
The proposed formalism aims for the description of the transition form factors, rather than the static properties of the baryons (masses, charge radii, etc.).

Inspired by previous studies based on quark models~\cite{Aznauryan07,JDiaz04a} we assume that the resonance $N(1440)$ can be interpreted as the first radial excitation of the nucleon, and use the properties of our model to calculate the $\gamma^\ast N \to N(1440)$ transition amplitudes and transition form factors.
Our calculations require no additional parameters apart from the ones used in the study of the structure of the nucleon~\cite{N1440}.

From the comparison with the JLab data~\cite{CLAS09,CLAS12,CLAS16a,Mokeev23a}, we conclude  that our model calculations are in good agreement with the measured data, above $Q^2=2$ GeV$^2$, supporting the assumption that $N(1440)$ can be in fact the first radial excitation of the nucleon and it is dominated by the effects of the valence quarks.
Our model calculations are, however, unable to provide an accurate description of the data in the region $Q^2=0$--1.5 GeV$^2$.
The gap between our calculations and the data at low $Q^2$ may be a consequence of our approximations for the Roper wave function, but can also be an indication that additional interactions besides the photon-quark interactions are necessary to describe the $\gamma^\ast N \to N(1440)$ transition.
The missing effects may be the contributions from baryon-mesons states, also called meson cloud contributions~\cite{Aznauryan12a,PPNP2024,NSTAR2017,Burkert04,Tiator04,Drechsel07}.
This interpretation suggests then a hybrid structure, where the quark substructure is revealed at large $Q^2$, and a molecular baryon-meson structure emerges in the low-$Q^2$ region.

Motivated by the success of the calculations for the $\gamma^\ast N \to N(1440)$ transition we use the covariant spectator quark model to calculate also the $\gamma^\ast N \to N^\ast$ transition amplitudes, where $N^\ast$ is the second radial excitation of the nucleon, at large $Q^2$~\cite{N1710}.
We apply the methodology implemented to the Roper, also to the study of the Roper version of the $\Delta(1232)$, associated with the radial excitation of the $\Delta(1232)\frac{3}{2}^+$, the $\Delta(1600)\frac{3}{2}^+$~\cite{Delta1600}.
Our calculations will be compared with the recent measurements of the $\gamma^\ast N \to \Delta(1600)$ amplitudes from JLab~\cite{Mokeev23a}.
Our studies of the electromagnetic transitions between the nucleon and the first and second radial excitations of the nucleon and on the $\gamma^\ast N \to \Delta(1600)$ transition will be discussed in the present article.

We also studied the $\gamma^\ast N \to N(1440)$ transition using a holographic model derived from AdS/QCD in leading order approximation, taking into account exclusively the contributions of the valence quarks~\cite{Roper-AdS1,Roper-AdS2}.
We conclude at the end that holographic estimates are similar to the estimates of the covariant spectator quark model.

We complement the discussion of the model calculations based on valence quark degrees of freedom with a discussion of the analytic properties of the $\gamma^\ast N \to N(1440)$ transition amplitudes near the photon point~\cite{Devenish76,Tiator16,LowQ2param}, and with the discussion of the relevance in taking into account the contributions associated with the meson cloud dressing of the baryon cores.

The present article is organized as follows:
In the next section we introduce the formalism associated with the $\gamma^\ast N \to N(1440)$ transition.
The theoretical frameworks used in the article: covariant spectator quark model and light-front holography are discussed in Sec.~\ref{sec-theory}.
Numerical calculations for the  $\gamma^\ast N \to N(1440)$ transition amplitudes, transition form factors, and a discussion about the parametrizations of the data at low-$Q^2$, are presented in Sec.~\ref{secResults}.
The $\gamma^\ast N \to \Delta(1600)$ transition is discussed in Sec.~\ref{sec-D1600}.
We finish with a general discussion of the present status of our understanding of the structure of the Roper, in Sec.~\ref{sec-Discussion}, and with the outlook and conclusions (Sec.~\ref{sec-Conclusions}).

\section{Electromagnetic transition between
  the nucleon and the Roper resonance \label{secDefinitions}}

We introduce now the formalism used in the study of the $\gamma^\ast N \to N^\ast$ transitions, and the $\gamma^\ast N \to N(1440)$ transition in particular.
In the following, we use $M$ for the mass of the nucleon, and $M_R$ for the mass of the nucleon excitations, respectively.

The study of the electromagnetic transition between the nucleon (momentum $P_N$) and the nucleon excitation $N^\ast$ (momentum $P_R$) requires the knowledge of the transfer momentum $q=P_R -P_N$ and the spin-parity ($J^P$) quantum numbers of the $N^\ast$ state.
The $\gamma^ \ast N \to N^\ast$ transition amplitudes are measured in electron nucleon scattering when $Q^2= -q^2 \ge 0$.

In the case of the Roper ($J^P=\frac{1}{2}^+$), the transition current takes the form
\ba
J^\mu = F_1 (Q^2)\left( \gamma^\mu - \frac{\slashed{q} q^\mu}{q^2}  \right) +
F_2 (Q^2) \frac{i \sigma^{\mu \nu} q_\nu }{M_R + M},
\label{eqCurrent}
\ea
where $\gamma^\mu$ is the Dirac matrix, $\sigma^{\mu \nu} \equiv \frac{i}{2}[\gamma^\mu, \gamma^\nu]$, and $F_1(Q^2)$ and $F_2(Q^2)$  define the Dirac and Pauli form factors, respectively.

The representation (\ref{eqCurrent}) requires that $F_1 (Q^2) \simeq Q^2  F_1^{\prime} (0)$ near $Q^2=0$, leading to $F_1 (0) =0$.
Alternative representations for the current (\ref{eqCurrent}) have been proposed.
See Refs.~\cite{PPNP2024,Aznauryan12a,Devenish76} for more detailed discussions.

For the definition of the transition amplitudes it is convenient to use the kinematics of the $N^\ast$ rest frame, in terms of the photon three-momentum ${\bf q}$
\ba 
P_N = (E,0,0,- |{\bf q}|), \hspace{.5cm} P_R= (M_R,0,0,0), \hspace{.5cm}
q= (\omega, 0,0, |{\bf q}|),
\ea
where
\ba
|{\bf q}| = \frac{\sqrt{Q_+^2 Q_-^2}}{2 M_R}, \hspace{1.cm} 
Q_\pm^2 = (M_R \pm M)^2 + Q^2, \\
E= \frac{M_R^2 + M^2 + Q^2}{2M_R},  \hspace{1.cm}
\omega = \frac{M_R^2- M^2- Q^2}{2 M_R}.
\ea
We will also use the fine-structure constant $\alpha \simeq 1/137$, and
\ba
K = \frac{M_R^2- M^2}{2 M_R}, \hspace{1.cm}  \tau = \frac{Q^2}{(M_R +M)^2}.
\ea

The transverse ($A_{1/2}$) and longitudinal ($S_{1/2}$) amplitudes associated with the $\gamma^\ast N \to N(1440)$ transition, also known as helicity amplitudes, can now be written in terms of the transition form factors, as 
\ba
& &
A_{1/2} (Q^2)= 
{\cal R}
\left[ F_1 (Q^2) + F_2  (Q^2) \right], 
\label{eqAmpA12} \\
& &
S_{1/2} (Q^2) =
\frac{{\cal R}}{\sqrt{2}}
|{\bf q}| \frac{M_R + M}{Q^2}
\left[ F_1 (Q^2) - \tau F_2 (Q^2)\right],
\label{eqAmpS12}
\ea
where
\ba
   {\cal R}= \sqrt{\frac{2 \pi \alpha}{K}} \sqrt{\frac{Q_-^2}{M M_R}}.
\label{eqCalR}
\ea

For the discussion of the properties of the $\gamma^\ast N \to N(1440)$ transition, it is worth mentioning that the form factors $F_1$ and $F_2$ are independent functions, while the helicity amplitudes are linear combinations of the form factors.
In Sec.~\ref{sec-Amp-param}, we comment on empirical parametrizations of the $\gamma^\ast N \to N(1440)$ transition amplitudes.

The transition form factors, $F_1$ and $F_2$, and the  helicity amplitudes $A_{1/2}$ and $S_{1/2}$, are well-defined in the region $|{\bf q}| \ge 0$, corresponding to the $Q^2  \ge -(M_R-M)^2$ region.
The point $|{\bf q}| = 0$, equivalent to  $Q^2 = -(M_R-M)^2$, is called pseudothreshold~\cite{Devenish76}.
The region $-(M_R-M)^2 \le Q^2 < 0$ cannot be accessed by 
electron nucleon scattering~\cite{Drechsel07,Tiator16}.
Near the pseudothreshold the helicity amplitudes
must have a well-defined dependence on $|{\bf q}|$~\cite{PPNP2024,Tiator16,LowQ2param}.
The properties of the $\gamma^\ast N \to N(1440)$ amplitudes in the low-$Q^2$ region are discussed in Sec.~\ref{sec-lowQ2}.

%%%%     Sec_Theory

\section{Theoretical models for the nucleon and $N^\ast$ electromagnetic structure \label{sec-theory}}

We discuss now some theoretical frameworks that have been used in the study of the $\gamma^\ast N \to N\left(\frac{1}{2}^+ \right)$  transitions.

\subsection{Covariant spectator quark model \label{secCSQM}}

The covariant spectator quark model is based on the covariant spectator theory~\cite{Gross69a,Gross99}.
In the formalism the baryons are treated, in the first approximation, as a system of three quarks.
The development of the formalism was more motivated by the description of the internal structure of the baryons, rather than the prediction of the baryon spectrum and the static properties of the baryons~\cite{NSTAR,PPNP2024,Nucleon,Omega,NSTAR2017}.

The spin-flavor structure of the wave functions of the baryons is derived from the SU(6) symmetry group.
In the formalism, however, the radial wave functions are not calculated by a wave equation with some complex and non-linear potential, as in most quark models.
The shape of the radial wave functions is instead estimated phenomenologically by empirical data or lattice data for some ground states~\cite{Nucleon,NSTAR2017}.

The electromagnetic transition current between baryon states is calculated using relativistic impulse approximation, where the photon interacts with an off-shell quark, while the two other quarks are on-shell spectators.
Within the covariant spectator theory, one can integrate over the momentum of the two on-shell quarks and reduce the three-quark wave function to a quark-diquark wave function, where the effective diquark has an averaged mass $m_D$~\cite{NSTAR,Nucleon,Nucleon2}.
The effective quark–diquark wave function $\Psi_B$ simulates the effect of confinement, and is consistent with the chiral symmetry when the mass of the quarks vanishes~\cite{NSTAR,Nucleon}.

The photon-quark coupling is determined by a quark current $j_q^\mu$ that can be decomposed in terms of Dirac ($j_1$) and Pauli ($j_2$) quark form factors, which encode effectively the gluon and quark–antiquark substructure of the constituent quarks.

The $\gamma^\ast B \to B'$ transition current can then be calculated in relativistic impulse approximation, using
\ba
J_{BB'}^\mu = \int_k \overline{\Psi}_{B'} (P_{B',}k) \; j_q^\mu \;  \Psi_{B} (P_B,k),
\hspace{.5cm}
j_q^\mu (Q^2) = j_1 \gamma^\mu + j_2 \frac{i \sigma^{\mu \nu}{q_\nu}}{2 M},
\label{eqCurrent2}
\ea
where $\Psi_{B}$ ($\Psi_{B'}$) is the wave function of the initial (final) state, $P_{B}$ ($P_{B'}$) is the initial (final) momentum, $k$ is the momentum of the diquark, and the integration sign represents the covariant integration in the diquark on-shell momentum.
For convenience, the quark Pauli form factor is defined in terms of the nucleon physical mass.
The previous relations can be used to calculate the $\gamma^\ast B \to B'$ transition form factors.

The quark Dirac and Pauli form factors can be decomposed into the functions $f_{i \ell} (Q^2)$, where $i=1,2$, and $\ell=0,\pm$, represent the strange quark component ($\ell =0$) and $+/-$ the isoscalar/isovector components.
These functions are parametrized using a vector meson dominance form based on a few vector meson mass poles ($\rho$, $\omega$, $\phi$,  etc.)~\cite{Omega,Octet}.

The covariant spectator quark model was originally developed for the nucleon electromagnetic structure~\cite{Nucleon}.
The studies for the nucleon established the parametrizations of the quark electromagnetic form factors $f_{i\pm}$ and the radial wave functions of the nucleon used in further applications of the model to the nucleon resonances.
Reviews about the calculations associated with nucleon resonances can be found in Refs.~\cite{PPNP2024,NSTAR2017}.

The covariant spectator quark model formalism has been generalized to the SU(3) flavor sector for the study of baryons with strange quarks (hyperons)~\cite{Omega,Octet,OctetDecuplet,OctetDecuplet2,Omega2,HyperonFF2}. 
Taking advantage of the quark form factor structure based on vector meson dominance and the form of the radial wave functions, the model has been extended to the lattice QCD regime (large pion masses)~\cite{Lattice,LatticeD,Octet}, to the timelike regime ($Q^2 < 0$)~\cite{Timelike,Timelike2,N1520TL,N1535TL}, and to the electromagnetic
and axial structure of baryons in the nuclear medium~\cite{Symmetry2025,GA-medium,GA-Octet,Octet3}.

The above discussion takes into account only the effects associated with the valence quark degrees of freedom.
There are, however, some processes, such as the meson exchanged between the different quarks inside the baryon, which cannot be reduced to processes associated with the dressing of a single quark.
Those processes can be regarded as a consequence of the meson exchanged between the different quarks inside the baryon, and can be classified as meson cloud corrections to the hadronic reactions~\cite{Octet,OctetDecuplet2,Hyperons-Dalitz}.

\subsection*{Radial excitations of the nucleon}

We focus now on the description of the radial excitations of the nucleon, including the Roper.
The nucleon and the Roper are characterized by the same spin-isospin structure.
The differences between the states come then from the differences between the radial wave functions.

The nucleon radial wave function $\psi_N$ is represented in the Hulthen form~\cite{Nucleon,N1440}
\ba
\psi_N (P,k) = \frac{N_0}{m_D \left( \beta_1 + \chi_{\ms N}\right)\left( \beta_2 + \chi_{\ms N}  \right)},
\hspace{.2cm}
\chi_{\ms N} = \frac{(M-m_D)^2-(P-k)^2}{M m_D},
\label{eq-psiN}
\ea
where $N_0$ is a normalization constant, $P$ is the nucleon momentum, and $\beta_1$ and $\beta_2$ are two dimensionless parameters that define the range of the radial wave functions in the momentum space.
Numerically, we consider $\beta_2 > \beta_1$, thus $\beta_1$ regulates the long-range structure, while $\beta_2$ regulates the short-range structure in the position space.
The form (\ref{eq-psiN}) ensures the expected falloff for the nucleon elastic form factors at large $Q^2$: $G_E$, $G_M$ $ \propto \;  1/Q^4$.

The Roper radial wave function, labeled here as $\psi_{N1}$ can be defined in the form
\ba
\psi_{N1} (P,k) =
\frac{N_1}{m_D \left( \beta_1 + \chi_{\ms N1}\right)\left( \beta_2 + \chi_{\ms N1}  \right)}
\frac{\beta_3 - \chi_{\ms N1}}{\beta_1 + \chi_{\ms N1}},
\label{eq-psiR}
\ea
where $ \chi_{\ms N1}$ is now defined in terms of the mass of the Roper,  $N_1$ is a new normalization constant, and $\beta_3$ is a new range parameter.

The interpretation of the new state as a radial excitation of the nucleon demands the orthogonality between the nucleon and Roper wave functions. 
The orthogonality condition can be expressed as $\int_k \Psi_{N1}^\dagger(\bar P,k) \Psi_N (\bar P,k) =0$, where $\bar P = \sfrac{1}{2}(P_N + P_R)$ in the limit where the difference of masses $M$ and $M_R$ can be neglected.
Here we use $\Psi_{N}$ and $\Psi_{N1}$ to represent the nucleon and the Roper wave functions, respectively.

In the calculations, we consider, that the difference of masses can be neglected in first approximation, meaning that the photon energy at the Roper rest frame $\omega= \frac{M_R^2-M^2}{2M_R}$ can be approximated by the photon point, leading to
\ba
\left. \int_k  \psi_{N1} \psi_N \right|_{Q^2=0} =0.
\label{eqOthogonal}
\ea
We use this equation to determine the parameter $\beta_3$, obtaining then a parameter-free representation of the Roper wave function.

Considering the nucleon wave function determined in previous studies, the Roper wave function determined by the orthogonality condition, and the established parametrizations of the quark form factors $f_{i\pm}$, we can calculate the $\gamma^\ast N \to N(1440)$ transition form factors, using Eq.~(\ref{eqCurrent2}), without any parameter fitting~\cite{N1440,N1710}.
Our model calculations, based on the Eqs.~(\ref{eq-psiR}) and (\ref{eqOthogonal}), are discussed in Secs.~\ref{sec-Amps1} and \ref{sec-FF}.

It is worth noticing that the model calculations described above are approximated, due to the assumptions considered for the masses, and because we took into account only the contributions associated with the valence quarks.
We expect nevertheless that the calculations provide a good description of the data in the large $Q^2$ region.

In the future, we may improve the calculations of the form factors using the semirelativistic approximation, discussed in Ref.~\cite{SemiR} for the nucleon resonances $N(1520)$ and $N(1535)$.
The basic idea of the approximation is that the spin-flavor structure is more relevant than the numerical effect of the masses, and that a good estimate of the transition form factors can be obtained replacing $M$ and $M_R$ by the average mass of the baryons.

The methodology discussed above to determine the wave function associated with the first radial excitation of the nucleon can be extended to the second radial excitation of the nucleon.
In that case, we need to consider the orthogonality conditions for three pairs of states~\cite{N1710}.
Our calculations of the transition amplitudes associated with the second radial excitation of the nucleon are discussed in Sec.~\ref{sec-Roper2}

The procedures discussed for the study of the radial excitations of the nucleon ($J^P= \frac{1}{2}^+$) can be applied to other $N^\ast$ states.
Of interest can be the first radial excitation of the first nucleon resonance, the $\Delta(1232)$ (isospin $I= \frac{3}{2}$ and $J^P = \frac{3}{2}^+$),
identified as the resonance  $\Delta(1600)$~\cite{Delta1600}.
The model calculations associated with the $\gamma^\ast N \to \Delta(1600)$ transition amplitudes and the comparison with the data are presented in Sec.~\ref{sec-D1600}.

\subsection{Light-Front holography \label{secLFH}}

In recent years it was shown that the combination of string theory of gravity in anti-de-Sitter (AdS) space can provide a description of conformal field theories (CFT) with strong couplings, like QCD in the confining regime~\cite{Brodsky15a,Maldacena98,Witten98}.
The application to QCD is possible due to the correspondence between the results from AdS/CFT and light-front dynamics based on a Hamiltonian with a confining mechanism of QCD~\cite{Brodsky15a,Teramond12a,Teramond12b,Gutsche13a}.
The framework is also known as AdS/QCD, light-front holography or holographic QCD.

Although the formalism is derived from the semiclassical approximation of QCD, it can be used to describe many properties of the hadronic systems due to its simplicity.
Of particular interest is the application to the structure of the baryons and to the electromagnetic transitions between baryon states.
We consider here top-down approaches to AdS/QCD, where the connection with light-front can be used to expand the wave functions in terms of states with a well defined number of constituents (quarks and gluons)~\cite{Brodsky15a}.
In these conditions some properties of the baryons can be described taking into account the first terms of the expansion using a small number of parameters~\cite{Brodsky15a,Teramond12a,Gutsche13a}.

Light-front holography can be applied to the study of the electromagnetic structure of the baryons, noting that the three-quark systems can be regarded as two-body systems with an active quark and a spectator quark pair.
In the calculations, the masses and wave functions are determined by the light-front wave equations~\cite{Brodsky15a}.
The contributions to a physical process can be interpreted and decomposed in terms of Fock states with well defined degrees of freedom, from the lowest order to higher orders.
The lowest order corresponds then to the system of three-quarks ($qqq$), while higher orders can have contributions associated with 4 constituents, as $(qqq)g$, where $g$ is a gluon, 5 constituents, including a $(q \bar q)$ excitation, as in $(qqq)(q \bar q)$, and so on~\cite{Brodsky15a}.

We consider in particular the formalism from Gutsche {\it et.al}~\cite{Gutsche13a}, according to the elastic form factors of the nucleon, and the nucleon to the Roper transition form factors can be determined in the lowest order by three independent couplings and an energy scale parameter.

In our work we use the elastic form factor data of the nucleon (proton and neutron) to calibrate the model and then to make predictions for the $\gamma^\ast N \to N(1440)$ transition form factors~\cite{Roper-AdS1}.
We consider only data above $Q^2=1.5$ GeV$^2$, in order to avoid the contamination from meson cloud effects.
We expect the predictions to be accurate in the large $Q^2$ region, above $Q^2 =2$ GeV$^2$.

Holographic calculations of the $\gamma^\ast N \to N(1440)$ transition form factors are presented and discussed in Sec.~\ref{sec-HolQCD}.

%\newpage

\section{Experimental results for transition amplitudes and transition form factors
\label{secResults}}

In this section, we present calculations of the $\gamma^\ast N \to N(1440)$ amplitudes and transition form factors based on the valence quark structure of the nucleon and the Roper, and compare the results with the experimental data.
We consider in particular data from JLab/CLAS, in the range $Q^2=0.3$--4.6 GeV$^2$, associated with one pion~\cite{CLAS09} and two pion production~\cite{CLAS12,CLAS16a,Mokeev23a} for the amplitudes $A_{1/2}$ and $S_{1/2}$.
The data for the form factors $F_1$ and $F_2$ can be extracted from the data for the amplitudes $A_{1/2}$ and $S_{1/2}$.

\subsection{Helicity amplitudes \label{sec-Amps1}}

\begin{figure}[t]
  \vspace{.3cm}
\includegraphics[width=2.35in]{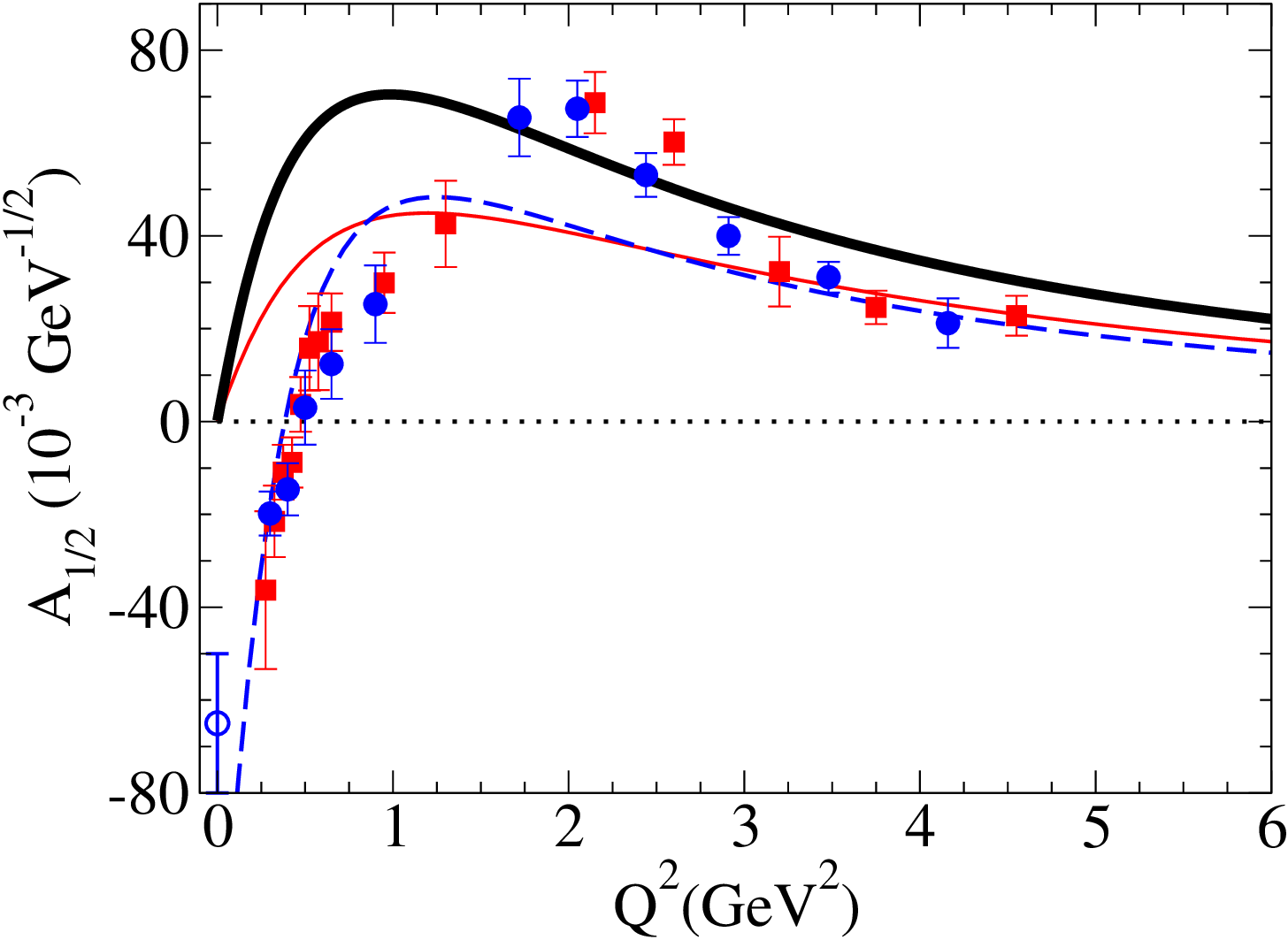} \hspace{.1cm}
\includegraphics[width=2.35in]{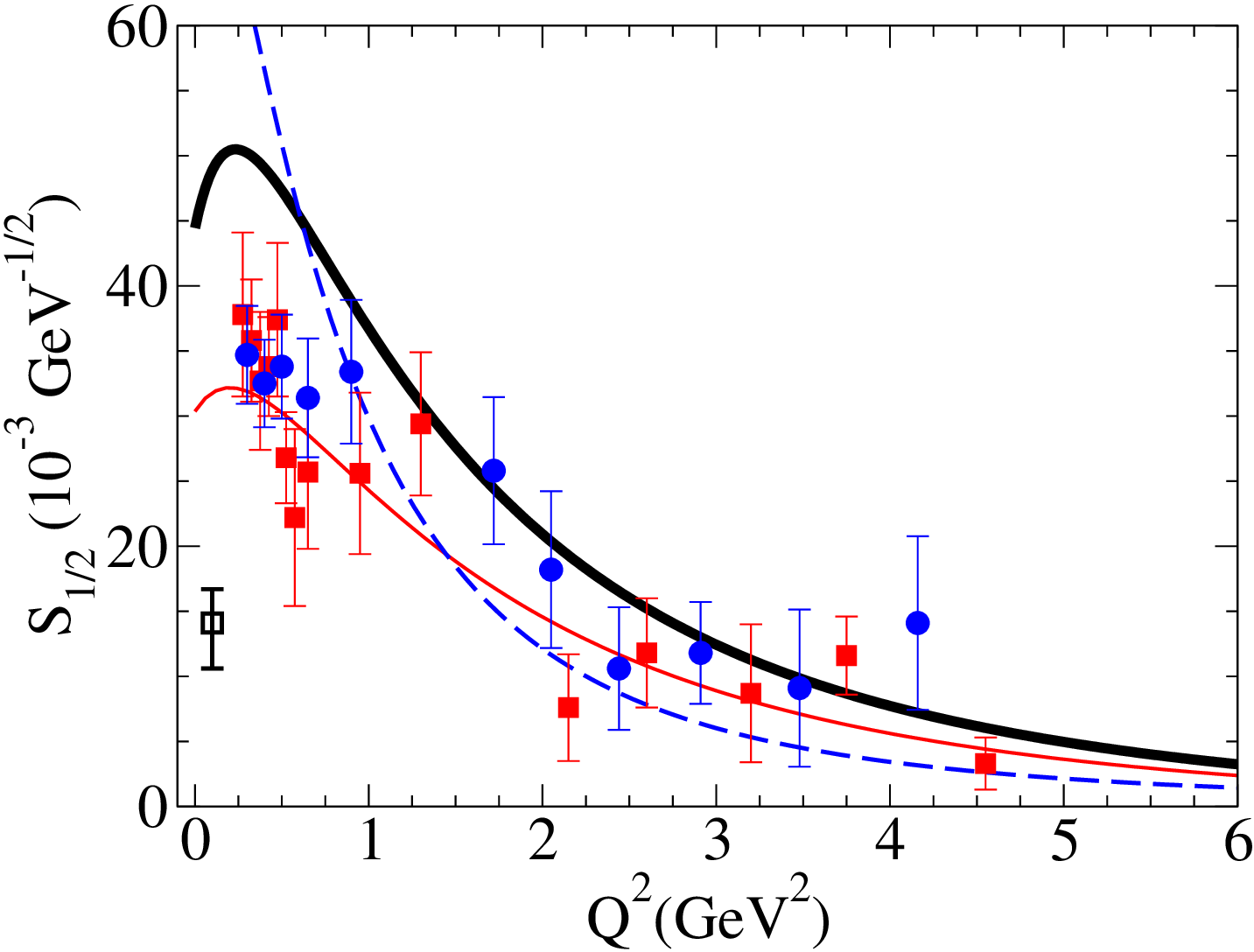} 
\caption{\footnotesize
  $\gamma^\ast N \to N(1440)$ helicity amplitudes.
  Calculations from covariant spectator quark model~\cite{N1440}
  for $M_R=1.440$ GeV (thick solid line) and $M_R= 1.600$ GeV (thin solid line), 
  and holographic model~\cite{Roper-AdS1} (dashed line).
  The data are from JLab/CLAS for one pion production (circles)~\cite{CLAS09}
  and two pion production (squares)~\cite{CLAS12,CLAS16a,Mokeev23a},
  and PDG 2024 (empty circle)~\cite{PDG2024}.
  For $S_{1/2}$ we include also a data point from MAMI at $Q^2=0.1$ GeV$^2$
  (empty square)~\cite{Stajner17}.
\label{fig-Roper-Amps1}}
\end{figure}

Our calculations of the $\gamma^\ast N \to N(1440)$ helicity amplitudes based on the covariant quark model, described in Sec.~\ref{secCSQM}, under the assumption that the $N(1440)$ is the first radial excitation of the nucleon,
are presented for the mass $M_R=1.440$ GeV
(thick solid line) in Fig.~\ref{fig-Roper-Amps1}.
No parameters are adjusted in the calculations, since the wave function of the Roper is determined by imposing the orthogonality with the nucleon wave function.
The relevant parameters of the model are determined in previous works in the study of nucleon electromagnetic form factors~\cite{Nucleon,N1440}.

%%%%    FIGURE 1

From Fig.~\ref{fig-Roper-Amps1}, one can conclude that the covariant spectator quark model for $M_R=1.440$ GeV
describes well the region above $Q^2=2$ GeV$^2$,
supporting the interpretation that the Roper is the first radial excitation of the nucleon.
In that respect, our calculations are in qualitative agreement with the predictions of Aznauryan from 2007 based on light-front quark model~\cite{Aznauryan07}, published before the measurements in 2009 for $Q^2 > 1$ GeV$^2$~\cite{Aznauryan12a,CLAS09}.

To study the sensibility to the resonance mass, we also calculated the transition amplitudes for the mass $M_R=1.600$ GeV (thin solid line).
From the comparison, one can conclude that the estimates are also close to the large-$Q^2$ data.
The direct comparison with amplitudes for $M_R=1.600$ GeV must be taken, however, with care, since the value of the mass enhances the amplitudes below $Q^2=1$ GeV$^2$, due to the amplification of the factor ${\cal R}$ from Eq.~(\ref{eqCalR}) for values above the Roper physical mass.

In the figure, we include as well a calculation from a light-front holographic model (dashed line), also based on the three-quark structure of the baryons.
Holographic calculations will be discussed in more detail later.

The models discussed here assume the dominance of the quark degrees of freedom.
It is not surprising then, to observe some deviations from the data in the low-$Q^2$ region, more specifically below $Q^2=1.5$ GeV$^2$.

In the low-$Q^2$ region it is expected that the contributions associated with quark-antiquark pairs and/or meson cloud excitations of the baryon bare cores are significant, as discussed commonly in the literature.
For a discussion on this subject, we recommend Refs.~\cite{Aznauryan12a,PPNP2024,Burkert19a,Burkert04,Tiator04}.
Chiral effective field theory and baryon-meson coupled-channel models improve in general the description of the low-$Q^2$ region~\cite{Wang24,Doring25a,Bauer14a,Gelenava18a}.
Also, quark models that take into account the baryon-meson dressing can help the description of the low-$Q^2$ region~\cite{Tiator04,Golli09a}.

Of particular interest are dynamical coupled-channel reaction models, dynamical models for short, applied to the baryon-meson reactions and to the analysis of the photo- and electro-production of mesons on nucleons~\cite{PPNP2024,Burkert04}.
In these models the meson cloud dressing of the baryons is taken into account using the baryon-meson interactions consistently.
Of relevance to the present discussion are the Sato-Lee/EBAC model~\cite{JDiaz09a,Suzuki10a,Matsuyama06a} and the ANL-Osaka model~\cite{Kamano13a,Nakamura15a,Kamano16a}.
Dynamical models are not formulated in terms of the quark degrees of freedom, but can be used to provide information about the structure of the baryon bare cores when the contribution from the meson cloud dressing is subtracted~\cite{PPNP2024,Burkert19a,Burkert04,JDiaz09a,Suzuki10a}.

A detailed discussion of calculations of the $\gamma^\ast N \to N(1440)$ amplitudes and form factors based on different quark model frameworks, models that take into account the baryon-meson contributions, and hybrid models can be found in Sec.~6.1 from Ref.~\cite{PPNP2024}.
The main conclusion from the comparison between models and data is that, in general, models where the valence quarks are the dominant component describe well the large-$Q^2$ region, including light-front quark models~\cite{Aznauryan07,JDiaz04a,Aznauryan12b,Obukhovsky14a}, the Dyson-Schwinger formalism~\cite{Segovia15a,Segovia16a}, and others.
Moreover, models based solely on the baryon-meson structure are not compatible with the observed falloff of the amplitudes and form factors~\cite{Burkert19a}.

\subsubsection{Remarks on the parametrizations of
  the $\gamma^\ast N \to  N(1440)$ amplitudes  \label{sec-Amp-param}}

Before discussing the numerical calculations for the $\gamma^\ast N \to  N(1440)$ form factors, it is important to discuss parametrizations of the $\gamma^\ast N \to  N(1440)$ amplitudes.
Parametrizations of the $\gamma^\ast N \to  N(1440)$ transition amplitudes are useful for studying scattering of electrons, photons and mesons with nucleons and nucleon resonances~\cite{PPNP2024,Burkert04,Drechsel07}.

As mentioned already, the amplitudes $A_{1/2}$ and $S_{1/2}$ are not independent functions, and are linear combinations of the $F_1$ and $F_2$.
As a consequence there are constraints on the transverse and longitudinal amplitudes that must be taken into consideration in the parametrizations of the data.
These constraints are often ignored in the literature, but may be relevant in the low-$Q^2$ region~\cite{LowQ2param}.
The impact of the low-$Q^2$ constraints on the shape of the amplitudes near $Q^2=0$, and the uncertainty of the data are discussed in Sec.~\ref{sec-lowQ2}.

Empirical parametrizations of the data have been derived in different forms.
We mention here four parametrizations of interest: the MAID parametrizations~\cite{Drechsel07}, the Rational functions from Ref.~\cite{Eichmann18}, the JLab para-metrizations~\cite{Blin19a,JLab-website}, and the JLab-ST parametrizations from Ref.~\cite{LowQ2param}.

The MAID parametrizations are based on the MAID analysis of different experiments~\cite{Drechsel07,MAID2011,MAID-website}.
The expressions for the amplitudes are defined as a combination of polynomials and exponentials, and are valid for the region $Q^2=0$--5 GeV$^2$.
The MAID parametrizations for the Roper are accurate for the amplitudes, but are not so appropriate for the form factors (see Sec.~6.1 of Ref.~\cite{PPNP2024}).

The Rational parametrizations from Ref.~\cite{Eichmann18} use ratios between polynomials, and are compatible with the expected power law falloffs of amplitudes at large $Q^2$ ($A_{1/2}$, $S_{1/2} \; \propto \; 1/Q^3$), and with the low-$Q^2$ constraints.

The JLab parametrizations are empirical parametrizations based mainly on the JLab/CLAS data, also defined in terms of rational functions~\cite{Blin19a,JLab-website}, and are valid for the range $Q^2=0.5$--6 GeV$^2$.
The  JLab-ST parametrizations proposed in Ref.~\cite{LowQ2param} modify the JLab parametrizations in the low-$Q^2$ region, in order to take into account the constraints near the pseudothreshold.

\newpage

\subsubsection{Helicity amplitudes of the second radial excitation
of the nucleon \label{sec-Roper2}}

The methodology used to calculate the wave function of the first radial excitation of the nucleon can be extended to the second radial excitation of the nucleon, also a $N\left(\frac{1}{2}^+ \right)$ state.
The $N\left(\frac{1}{2}^+ \right)$ states are not all associated with the radial excitations of the nucleon, and can differ in the structure of the wave functions~\cite{PDG2024,PPNP2024,Capstick00}.

Our estimations of the transition amplitudes associated with the transition from the nucleon to the second radial of the nucleon are presented in Ref.~\cite{N1710}.
The interpretation that the state might be the $N(1710)$ state has been discarded based on the analysis of the structure of the different $N\left(\frac{1}{2}^+ \right)$ states.
The second radial excitation is nowadays tentatively identified as the state $N(1880)$.

\begin{figure}[t]
  \vspace{.3cm}
\includegraphics[width=2.35in]{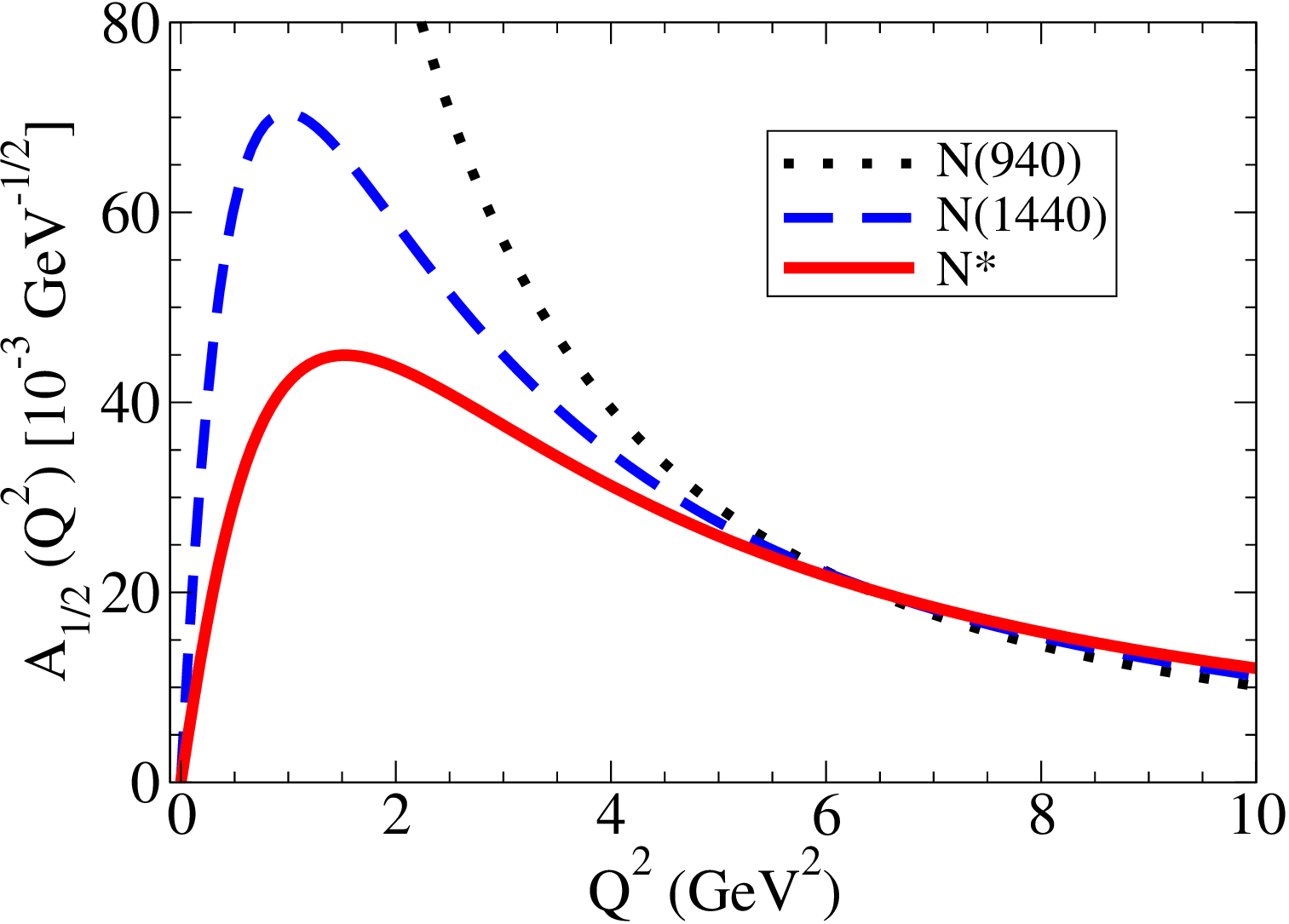} \hspace{.1cm}
\includegraphics[width=2.35in]{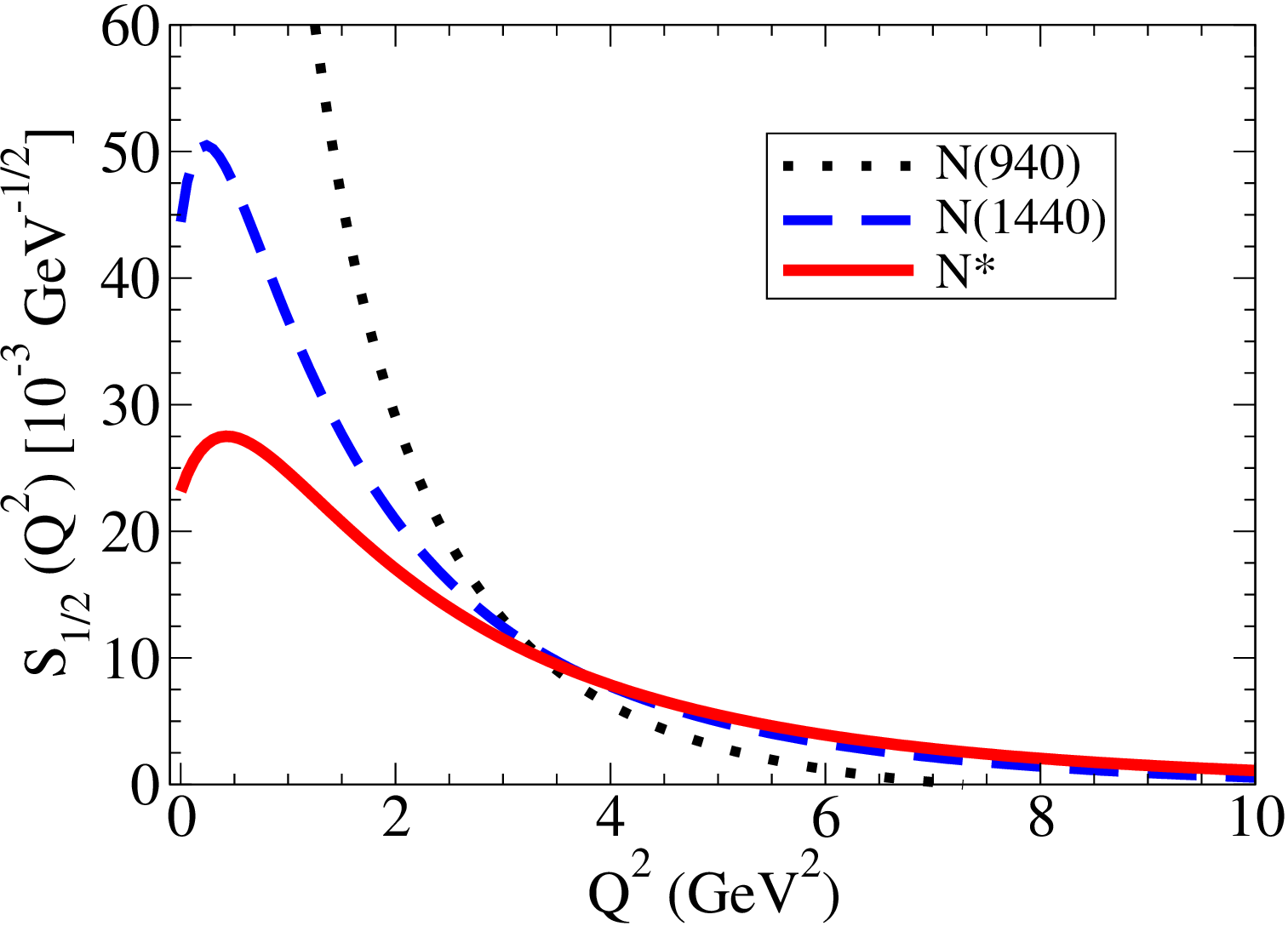} 
\caption{\footnotesize
  Calculations of the $\gamma^\ast N \to N^\ast$ helicity amplitudes
  for the first and second radial excitations
  of the nucleon.
  The equivalent amplitudes for the nucleon, $N(940)$,  are defined in the main text. 
\label{fig-Roper-Amps2}}
\end{figure}

The numerical calculations of the transition amplitudes associated with the second radial excitation of the nucleon (labeled as $N^\ast$) are presented in Fig.~\ref{fig-Roper-Amps2}.
As for the Roper, the model calculations are expected to be accurate for large $Q^2$, when the contributions from the meson cloud are reduced significantly.

In the figure, we also compare the results with the $\gamma^\ast N \to N(1440)$ amplitudes, and with the equivalent amplitudes for the nucleon.
The equivalent amplitudes for the nucleon are defined by $A_{1/2} \equiv \sqrt{2} \, {\cal R} \, G_M$, and $S_{1/2} \equiv {\cal R} \, \sqrt{\frac{1 + \tau}{\tau}} \, G_E$, where $G_M$ and $G_E$ are the proton elastic form factors, and ${\cal R}$ is defined by Eq.~(\ref{eqCalR}) in terms of the mass of the Roper.
The global factor $\sqrt{2}$ was included by convenience.

The nucleon equivalent amplitudes are used to emphasize the numerical similarity between the amplitudes of the Roper and the state $N^\ast$ identified tentatively as $N(1880)$.
Notice also how close the expressions for the transition amplitudes (\ref{eqAmpA12}) and (\ref{eqAmpS12}) are with the nucleon elastic form factors, defined by the combination of the Dirac and Pauli form factors for the nucleon $G_E \equiv F_1 - \tau F_2$ and $G_M \equiv F_1 + F_2$.

From the observation of Fig.~\ref{fig-Roper-Amps2}, we can notice the similarity of the amplitudes $A_{1/2}$ and $S_{1/2}$ for the three cases, above $Q^2=4$ GeV$^2$.
The exception is the equivalent amplitudes $S_{1/2}$ for the nucleon.
The amplitude appears to have a zero near $Q^2 \simeq 7.0$ GeV$^2$, consistently with the measurements at JLab~\cite{Arrington07a,Puckett18a}.
One can interpret the approximated convergence for the amplitudes as a consequence of the correlations between the radial wave functions of the excited states and the nucleon radial wave function~\cite{N1710}.

The corollary of the  resemblance of the results for the amplitudes at large $Q^2$ is that we can use the $G_M$ data (known up to $Q^2 \simeq 30$ GeV$^2$) for the nucleon to predict the amplitudes $A_{1/2}$ for the Roper and $N(1880)$ state for extremely large values of $Q^2$
Future measurements at JLab-12 GeV upgrade may confirm or deny these predictions.

%%%     sec4.2

\subsection{Transition form factors \label{sec-FF}}

Although experimental studies of the $\gamma^\ast N \to N^\ast$ transitions are almost exclusively performed using helicity amplitudes [see Eqs.~(\ref{eqAmpA12}) and (\ref{eqAmpS12})], there are advantages in analyzing the data using the transition form factors.
The data associated with the transition form factors can be obtained by inverting Eqs.~(\ref{eqAmpA12}) and (\ref{eqAmpS12}).

\begin{figure}[t]
  \vspace{.3cm}
\includegraphics[width=2.35in]{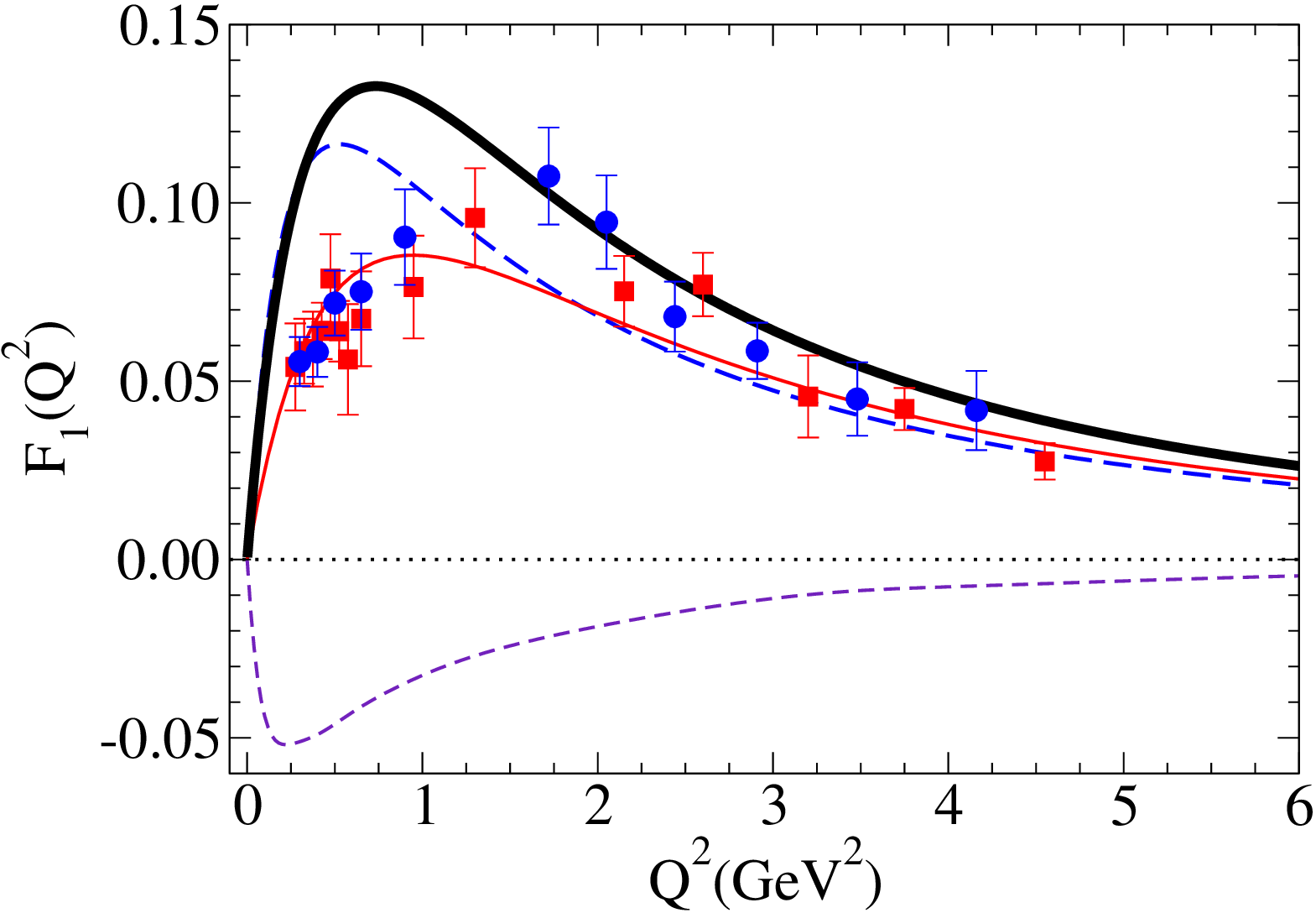} \hspace{.1cm}
\includegraphics[width=2.35in]{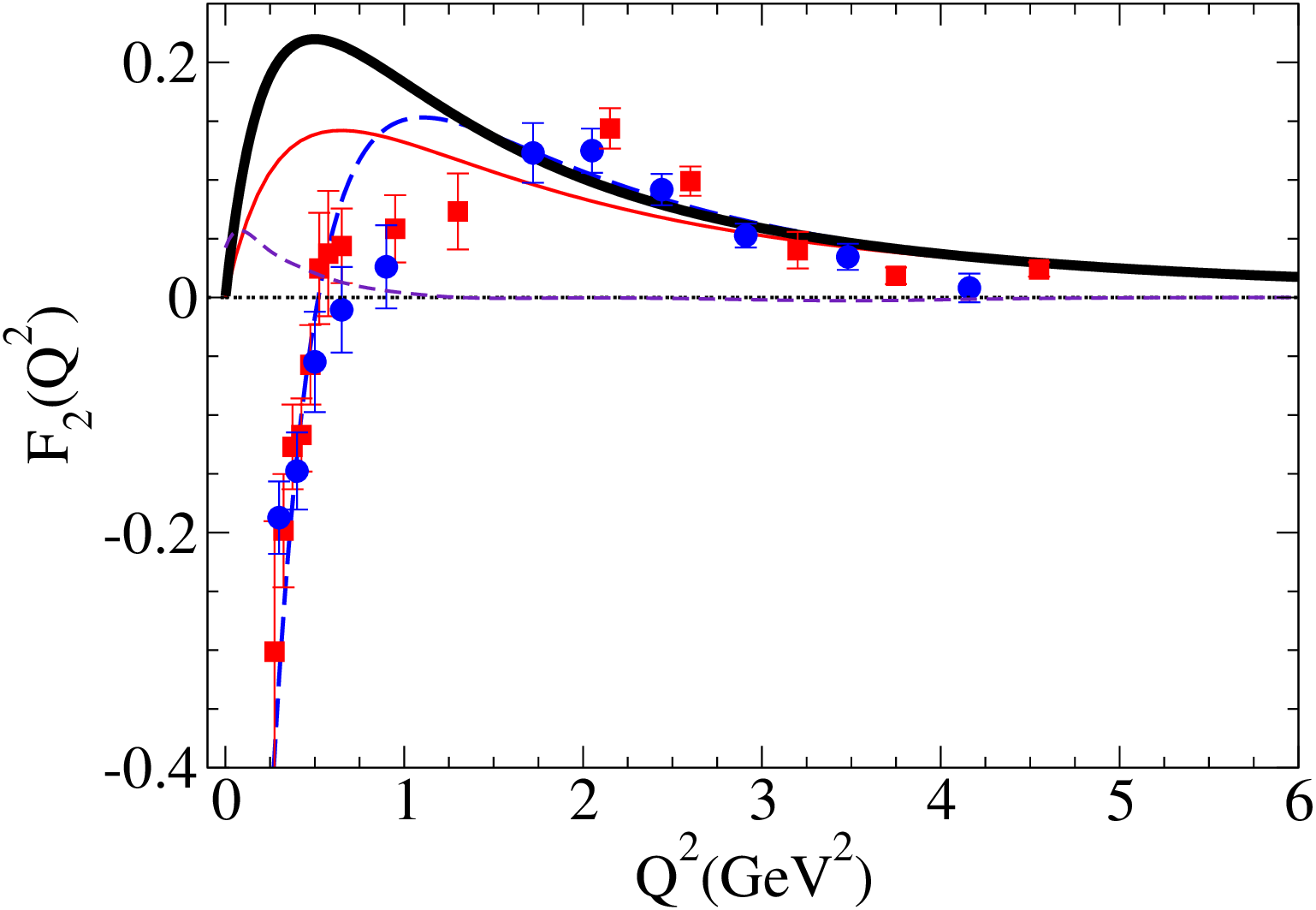} 
\caption{\footnotesize
  $\gamma^\ast N \to N(1440)$ transition form factors.
   Calculations from covariant spectator quark model~\cite{N1440}
   for $M_R=1.440$ GeV (thick solid line) and $M_R= 1.600$ GeV (thin solid line),
   and holographic model~\cite{Roper-AdS1} (dashed line).
 The data are from JLab/CLAS for one pion production (circles)~\cite{CLAS09}
  and two pion production (squares)~\cite{CLAS12,CLAS16a,Mokeev23a}.
\label{fig-Roper-FF1}}
\end{figure}

The covariant spectator quark model calculations for the $\gamma^\ast N \to N(1440)$ transition form factors, for the physical mass ($M_R-1.440$ GeV), are presented (thick solid line) in Fig.~\ref{fig-Roper-FF1}.
The quality of the description of the data for the Dirac and Pauli form factors became clear for large $Q^2$.
As noted also for the helicity amplitudes, there is a gap between the calculations and the data below $Q^2=1.5$ GeV$^2$, which may be interpreted as a consequence of the lack of meson cloud effects~\cite{PPNP2024,Burkert19a,N1440,N1440-proc}.

The form factors associated with the mass $M_R = 1.600$ GeV (thin solid line) are presented in Fig.~\ref{fig-Roper-FF1}.  
Also for this mass, we obtain a good agreement with the large-$Q^2$ data.

The comparison between models and data is cleaner when we consider form factors, because form factors are independent functions, and also because there is no need to include a multiplicative function ${\cal R}$, sensitive to the value of $M_R$ and to the low-$Q^2$ region.

%%%     FIGURE 2

Explicit calculations of the meson cloud contributions for the  $\gamma^\ast N \to N(1440)$ transition amplitudes have been performed by the ANL-Osaka group using a dynamical couple-channel model~\cite{Kamano13a,Nakamura15a,Kamano16a}.
The ANL-Osaka estimates\footnote{Transition form factor calculations based on dynamical coupled-channel models are complex functions. For simplicity, we include only the real part.} of the meson cloud contributions for the form factors (short-dashed line) are also included in Fig.~\ref{fig-Roper-FF1}.

Notice in Fig.~\ref{fig-Roper-FF1} that the meson cloud contributions to the $F_1$ form factor at low $Q^2$ are significant and negative.
When combined with the contributions from valence quarks, for $M_R=1.440$ GeV, we can then expect a reduction form factor below $Q^2=1$ GeV$^2$, consistent with the measured data.
The ANL-Osaka estimate for $F_1$ is close to the inferred estimate of the meson cloud contributions used in works based on the Dyson-Schwinger formalism~\cite{Burkert19a,Segovia16a}.
As for the form factor $F_2$, there are some discrepancies between the combined effects of valence quarks and meson cloud.
Alternative estimates of the inferred meson cloud contributions can be found in light-front quark models~\cite{CLAS16a,Aznauryan12b} and in some of our previous works~\cite{N1440,N1440-proc}.

Concerning the calculations based on the valence quark structure, we call the attention to model calculations that take into account the dependence of the quark mass on the momentum of the quarks, a consequence of the dynamical chiral symmetry breaking~\cite{Burkert19a}.
This mechanism takes into account the gluon dressing of quarks and interaction vertexes that lead to the  dynamical generation of the mass of the quarks~\cite{NSTAR,Burkert19a,Segovia15a}.
Examples of these calculations are the light-front quark model from Ref.~\cite{Aznauryan12b} and Dyson-Schwinger calculations~\cite{Segovia15a,Segovia16a}.

The covariant spectator quark model can also be applied to the calculation of the transition form factors in lattice QCD characterized by unphysical values for the pion mass.
In the calculations, we use the baryon masses ($M$ and $M_R$) and the vector meson masses, determined in the lattice QCD simulations, to define the quark current and radial wave functions.
Our calculations agree well with the available lattice QCD simulations for the $\gamma^\ast N \to N(1440)$ transition from Ref.~\cite{Lin08a} ($m_\pi \simeq 700$ MeV), below $Q^2=2$ GeV$^2$. 
These results support the assumption of the dominance of the valence quark contributions for simulations with large pion masses, when meson cloud effects are reduced significantly.

From the discussion of the model calculations of the $\gamma^\ast N \to N(1440)$ transition amplitudes and form factors, we conclude that the more successful descriptions of the data combine valence quark effects (dominant at large $Q^2$) and effective descriptions of the baryon-meson states (meson cloud), fundamental at low $Q^2$~\cite{PPNP2024,Burkert19a}.

\subsubsection{Holographic calculations of $\gamma^\ast N \to N(1440)$
  transition form factors  \label{sec-HolQCD}}

In Fig.~\ref{fig-Roper-FF1}, we also include the calculations of the holographic QCD model (long dashed line) presented in Fig.~\ref{fig-Roper-Amps1}.
From the observation of the two figures, one can conclude that the holographic model also provides a good qualitative description of the large $Q^2$ data.

We focus now on the holographic calculations of the transition form factors.
In the first studies of the  $\gamma^\ast N \to N(1440)$ transition within AdS/QCD, Teramond and Brodsky derived a parameter free expression for the Dirac form factor $F_1$ in terms of the physical masses of the $\rho$ meson and their first excitations (parameter free calculation)~\cite{Teramond12a,Teramond12b}.
The calculation assumes the dominance of the quark degrees of freedom, and that $N(1440)$ is the first radial excitation of the nucleon.

\begin{figure}[t]
  \vspace{.3cm}
\includegraphics[width=2.35in]{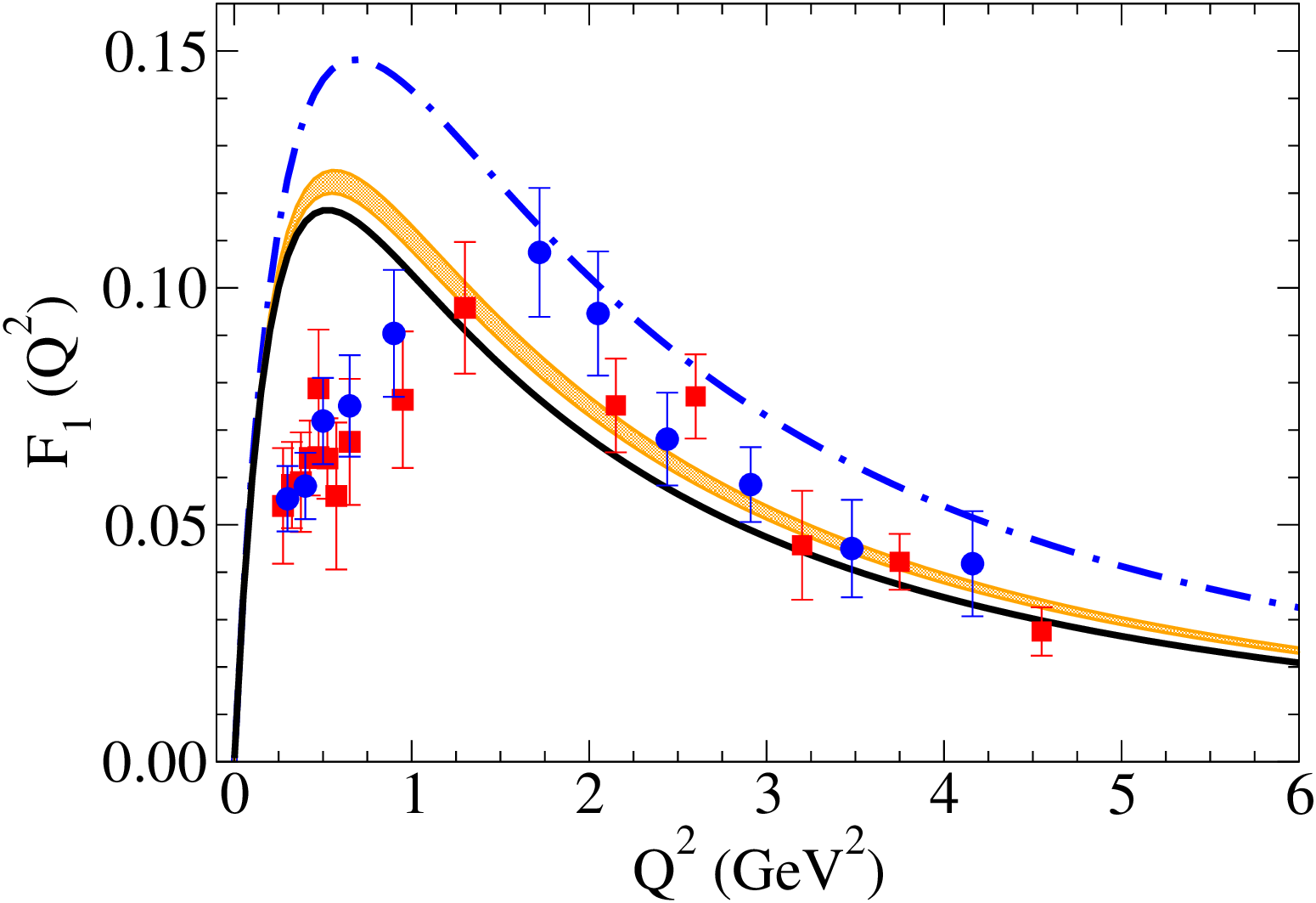} \hspace{.1cm}
\includegraphics[width=2.35in]{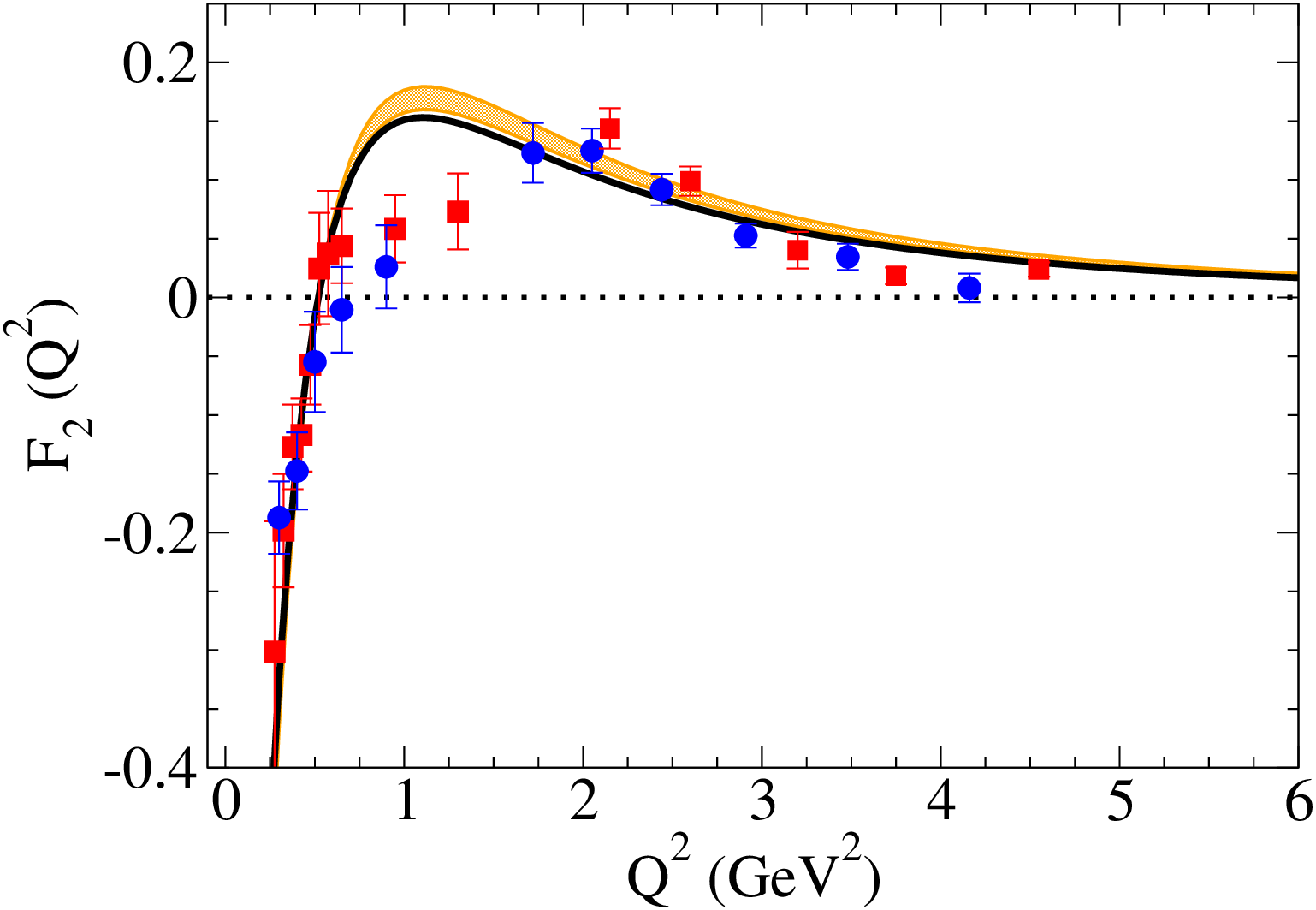} 
\caption{\footnotesize
  Holographic calculations of the $\gamma^\ast N \to N(1440)$ form factors,
  in leading order (nucleon and Roper described as $(qqq)$ states).
  Calculations from Teramond and Brodsky (dashed dotted line)~\cite{Teramond12a}
  and from Ref.~\cite{Roper-AdS1} (variation band).
  The solid line correspond to the holographic parametrization from Ref.~\cite{Roper-AdS2}.
  The data are described in Fig.~\ref{fig-Roper-FF1}.
\label{fig-Roper-FF2}}
\end{figure}

The result of Teramond and Brodsky (dashed doted line) is presented in Fig.~\ref{fig-Roper-FF2}, in comparison with the form factor data.
This work motivated the use of holographic models to calculate the $\gamma^\ast N \to N(1440)$ transition form factors, taking into account additional contributions to the $(qqq)$ structure~\cite{Gutsche13a,Gutsche18a}. 
In these calculations the gluon and $q \bar q$ excitations are also taken into account, but the effective weight of the higher Fock states was, in general, adjusted to some experimental data (transition amplitudes and hadron masses).

In Fig.~\ref{fig-Roper-FF2}, we also include our calculation of the transition form factors~\cite{Roper-AdS1} based on the formalism from Ref.~\cite{Gutsche13a}, where the free parameters were adjusted by the proton and neutron  elastic form factor data, above $Q^2=1.5$ GeV$^2$,  to avoid the contamination of the $q\bar q$ and $(q\bar q)(q\bar q)$ states.
Our calculations are represented by a band of variation that takes into account the uncertainty of the parameters in the fit to the nucleon data.
In the same figure we also include a simplified parametrization of the transition form factors in terms of physical masses of hadrons, proposed in Ref.~\cite{Roper-AdS2}.

From the comparison of the calculations based on the three-quark structure for the nucleon and Roper, we can conclude that different descriptions provide similar results in the large-$Q^2$ region.
The calculation of Teramond and Brodsky overestimates the data for $F_1$, but the remaining estimates are compatible with the data above $Q^2=2$ GeV$^2$, the region where we expect the dominance of valence quark effects.

From the observation of Figs.~\ref{fig-Roper-Amps1} and \ref{fig-Roper-FF1}, where the central value of our holographic model~\cite{Roper-AdS1} is included (dashed line), we can conclude that the covariant spectator quark model and the holographic model provide very similar results for large $Q^2$ (above $Q^2=2$ GeV$^2$).
Our conclusion is then that the physics associated with the quark degrees of freedom is well represented in both formalisms.

A more detailed discussion of the light-front holographic calculations of the $N(1440)$ transition form factors, including higher Fock states, can be found in Sec.~6.1 of Ref.~\cite{PPNP2024}.

Our main conclusion about holographic QCD is that the formalism can be a very useful tool to calculate the contributions from the bare core (first approximation), but that corrections to the leading order should be taken with care.
The limitations of the formalism are a consequence of the semiclassical approximation from AdS/QCD, and on the approximated description of the $q \bar q$ structure of the meson cloud contributions~\cite{PPNP2024,GA-AdS}.

%%%%     sec4.3

\subsection{Transition amplitudes at low $Q^2$ \label{sec-lowQ2}}

We discuss now the  $\gamma^\ast N \to N(1440)$ amplitudes in the low-$Q^2$ region.

There are two main aspects to discuss: the quality of the data, and the impact of the analytic constraints near the pseudothreshold in the shape of the amplitudes near $Q^2=0$.
Recall that the amplitudes $A_{1/2}$ and $S_{1/2}$ are linear combinations of the independent functions $F_1$ and $F_2$.

At the moment, the  $\gamma^\ast N \to N(1440)$ helicity amplitudes are not well known experimentally in the low-$Q^2$ region.
Except for the MAMI measurement, there are no data between the photon point and $Q^2=0.3$ GeV$^2$.

Contrary to other well-established resonances, such as the $\Delta(1232)$, \mbox{$N(1520)$} and $N(1535)$~\cite{Siegert1,Siegert3,N1535-ST,N1520-ST1,N1520-ST2},
for which there are strong correlations between the
transverse amplitudes and the longitudinal amplitude $S_{1/2}$,
in the case of the $N(1440)$ resonance, there is no direct correlation
between the amplitudes near the pseudothreshold.
The main constraints on the amplitudes are their analytical form~\cite{LowQ2param} 
\ba
A_{1/2} \; \propto \; |{\bf q}|, \hspace{1.3cm} S_{1/2} \; \propto \; |{\bf q}|^2,
\label{eqAmpsPT}
\ea
when  $|{\bf q}| \to 0$, where $|{\bf q}|$ is the modulus of the photon  three-momentum at the resonance rest frame.

The condition  $|{\bf q}| = 0$, is equivalent to $Q^2= -(M_R-M)^2$, a point below the photon point, in the timelike region.
Although the pseudothreshold \mbox{$Q^2= -(M_R-M)^2$} $ \simeq 0.25$ GeV$^2$, is well below $Q^2=0$, the conditions (\ref{eqAmpsPT}) may still have an impact on the shape of the amplitudes $A_{1/2}$ and  $S_{1/2}$ near $Q^2=0$.

\begin{figure}[t]
  \vspace{.3cm}
\includegraphics[width=2.35in]{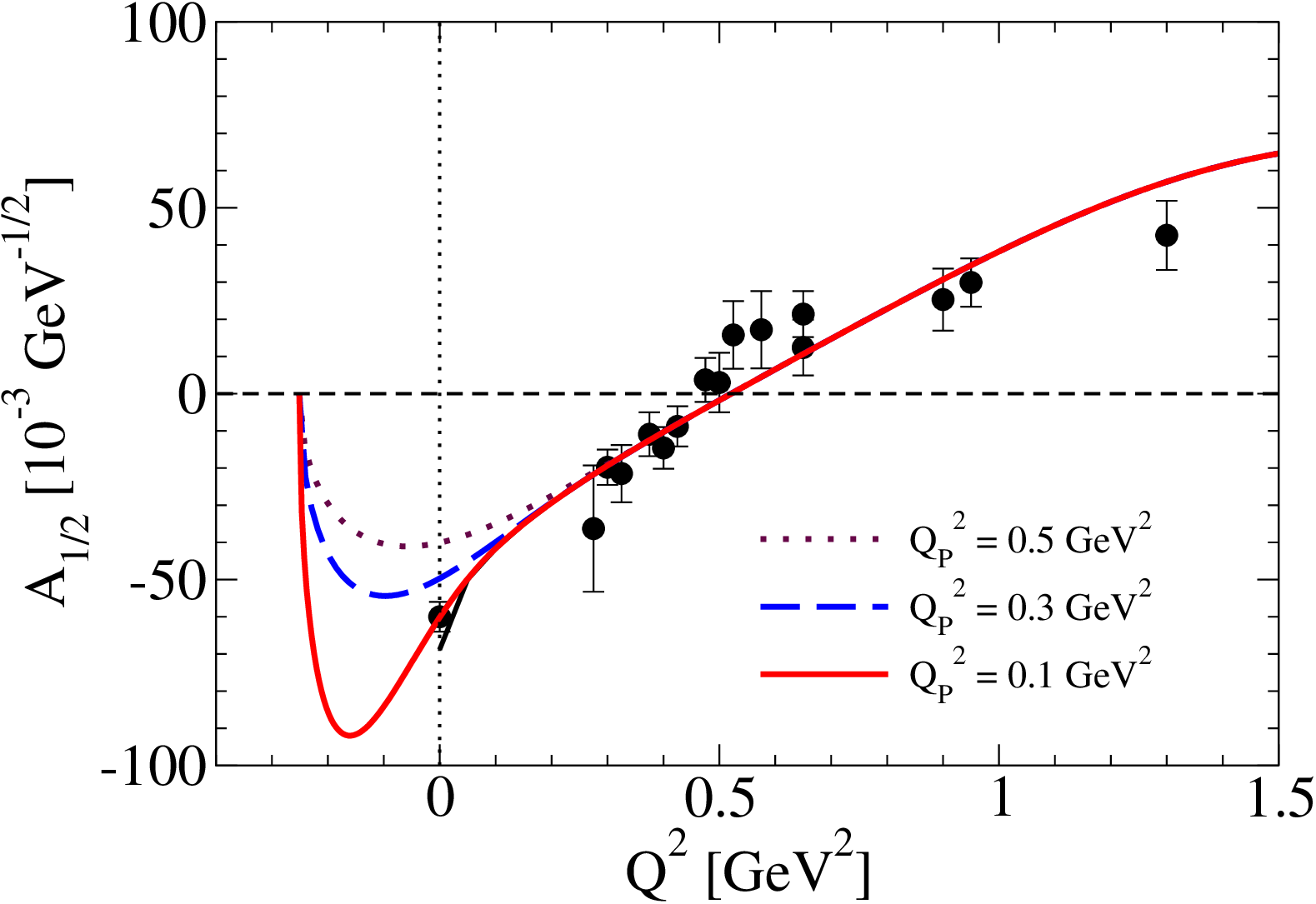}  \hspace{.15cm}
\includegraphics[width=2.35in]{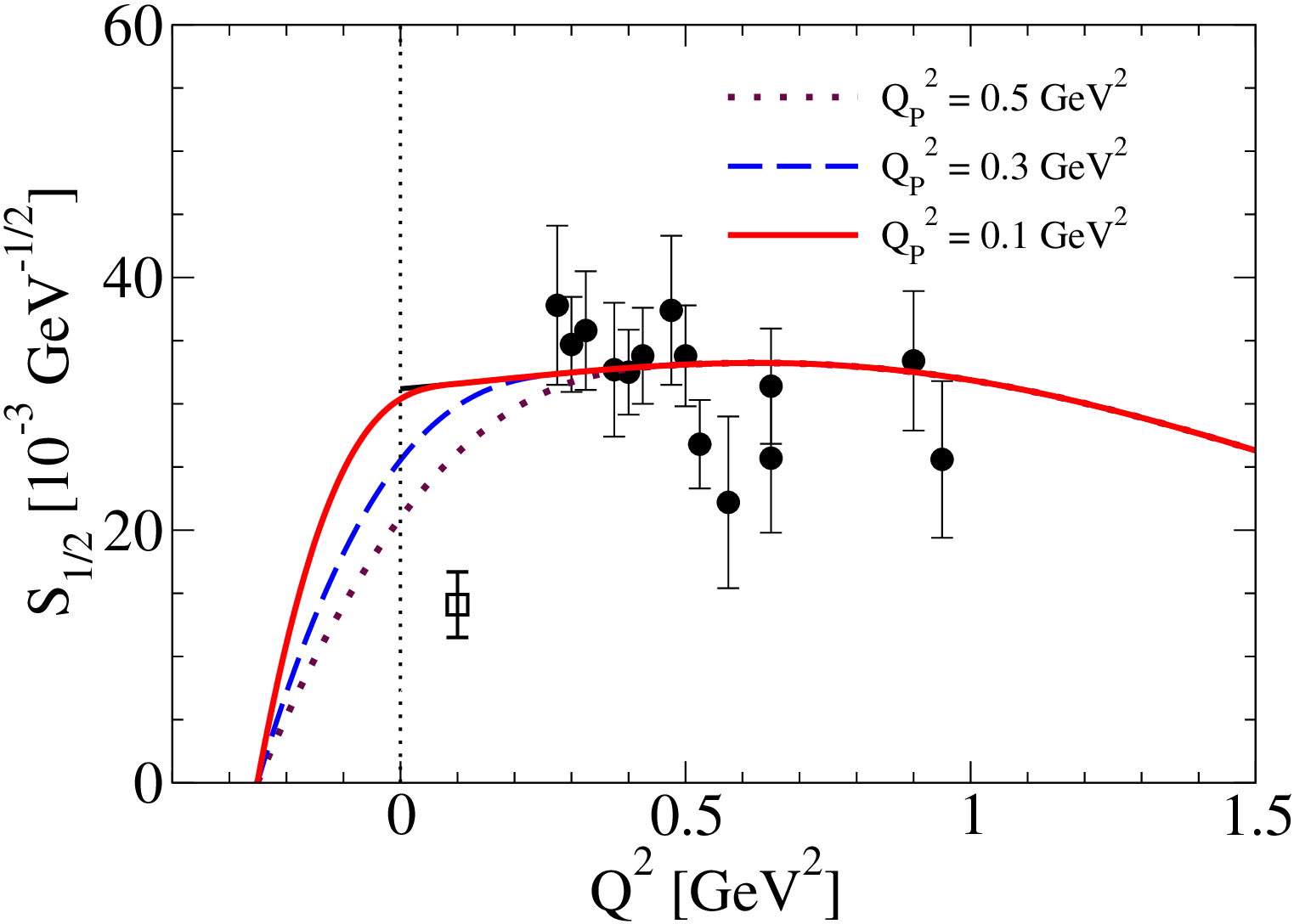} 
\caption{\footnotesize
  Parametrizations of the $\gamma^\ast N \to N(1440)$ amplitudes~\cite{LowQ2param}.
  The solid line represents the original JLab parametrization~\cite{Blin19a,JLab-website}.
  The data are from JLab/CLAS (circles)~\cite{CLAS09,CLAS12,CLAS16a} (one and two pion production), and MAMI (empty square) for $S_{1/2}$ ($Q^2=0.1$ GeV$^2$)~\cite{Stajner17}.
  The vertical dotted line represents the photon point ($Q^2=0$).
\label{fig-Roper-Amps3}}
\end{figure}

The effect of the relations (\ref{eqAmpsPT}) can be observed in Fig.~\ref{fig-Roper-Amps3}, where different analytic extensions of the JLab parametrizations below $Q^2=0.5$ GeV$^2$ are considered.
We labeled these parametrizations as JLab-ST, since they correspond to modifications of the JLab parametrizations, successful in the description of the large $Q^2$ data, in the low-$Q^2$ region, where Siegert's theorem must be fulfilled~\cite{Devenish76,Tiator16}.
We consider different analytical extensions starting at the point $Q_P^2$, labeled by the value of $Q_P^2$, down to $Q^2=0$ and to the pseudothreshold.
Above $Q_P^2$ the parametrizations are identical to the JLab parametrizations.
Below $Q_P^2$ the parametrizations represent smooth analytical continuations compatible with the conditions (\ref{eqAmpsPT}).
For a clear visualization, we consider $Q_P^2=0.1$, 0.3 and 0.5 GeV$^2$.

In Fig.~\ref{fig-Roper-Amps3}, we present also the amplitudes for negative values of $Q^2$ down to the pseudothreshold, in the timelike region, where we expect the amplitudes to vanish.
Notice in the figure, the change in the shape of the amplitudes near $Q^2=0$, according to the value $Q_P^2$.
Only the parametrizations $Q_P^2=0.1$ and 0.3 GeV$^2$ are compatible with the data for $A_{1/2}$.
The significant variation of the parametrizations with small variations in the value of  $Q_P^2$ shows how sensitive the parametrizations can change with variations of the data, and how important it is to make measurements of the transition amplitudes below $Q^2=0.3$ GeV$^2$.

Future measurements may help to understand if the data from MAMI ($Q^2=0.1$ GeV$^2$) is compatible with the JLab/CLAS data and with the pseudothreshold constraints.
Those measurements may also reduce the interval of variation of the amplitude $S_{1/2}$ near $Q^2=0$.

The kinematic region between the pseudothreshold and the photon point,  $-(M_R^2-M)^2 \le Q^2 < 0$, cannot be probed by spacelike photons, using electron nucleon scattering.
Under study at HADES and PANDA is, however, the possibility of measuring transition amplitudes and transition form factors in the timelike region for the first $N^\ast$ resonances~\cite{Timelike2,N1520TL,N1535TL,Hyperons-Dalitz,HADES2017,HADES2021}.   
Some progress has been done for the $\Delta(1232)$~\cite{PPNP2024,HADES2017}.
The $\Delta (1232) \to e^+ e^- N$ decay width has been measured and included in PDG~\cite{PDG2024}.

\section{Transition amplitudes of the first radial excitation of the $\Delta(1232)$
\label{sec-D1600}}

The $\Delta(1232)$ equivalent to the Roper state, interpreted as the first radial excitation, can be identified as the $\Delta(1600)$.
As most of the nucleon resonances of the third resonance region ($M_R > 1.55$ GeV) the properties of the $\gamma^\ast N \to \Delta(1600)$ transition are almost unknown, except for the photocouplings.

This picture changed recently with the measurements of the $\gamma^\ast N \to \Delta(1600)$ amplitudes in the range $Q^2=2.0$--4.5 GeV$^2$ at JLab/CLAS~\cite{Mokeev23a}. 
Before these measurements the signs of the amplitudes $A_{1/2}$ and $A_{3/2}$ were unknown.

Calculations of the valence quark contributions to the $\gamma^\ast N \to \Delta(1600)$ transition amplitudes and electromagnetic form factors ($G_E$ and $G_M$) were performed in 2010~\cite{Delta1600} using the methodology from the $\gamma^\ast N \to N(1440)$ transition.
We assumed that the $\Delta(1600)$ state is the first radial excitation of the $\Delta(1232)$, and that the wave functions were dominated by the valence quarks~\cite{NDelta,Delta1600}.
We use then the orthogonality to determine the radial wave function of the $\Delta(1600)$, since the spin-isospin structure is the same for both resonances~\cite{Delta1600}.
From the dominance of the S-state quark-diquark contributions, we concluded that $G_E \equiv 0$, and that the amplitudes are determined by
\ba
A_{1/2} (Q^2) = - R \, G_M(Q^2), \hspace{1cm}
A_{3/2} (Q^2) = - \sqrt{3}\, R \, G_M(Q^2), 
\ea
where $R =\frac{0.07158}{\sqrt{M}} \sqrt{\frac{(M_R-M)^2 + Q^2}{(M_R-M)^2}}$
[here $M_R$ is the mass of the $\Delta(1600)$].

The previous expressions can be used to calculate the transverse amplitudes, using the model calculations for $G_M$ and the sign of $G_M$ determined by the experimental data.
In Ref.~\cite{Delta1600}, we also estimate the magnitude of the meson cloud contributions, taking into account baryon-meson states associated with the channels $\pi N$, $\pi N(1440)$, $\pi \Delta(1232)$ and $\pi \Delta(1600)$.
We concluded that the meson cloud contributions to the magnetic form factor were in the range $|G_M^{\rm MC} (0)|=0.0$--1.3.

\begin{figure}[t]
  \vspace{.3cm}
\includegraphics[width=2.35in]{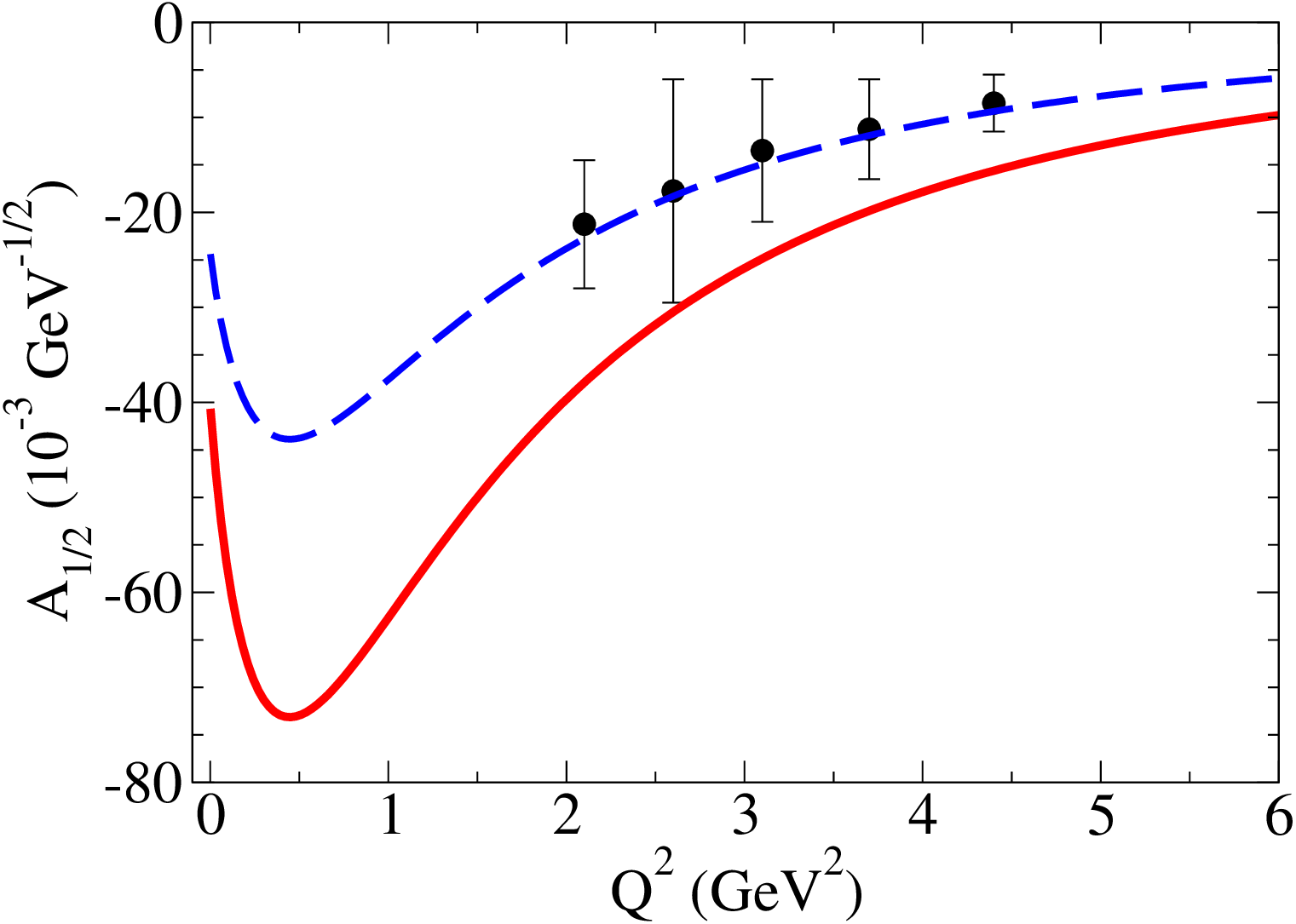} \hspace{.1cm}
\includegraphics[width=2.35in]{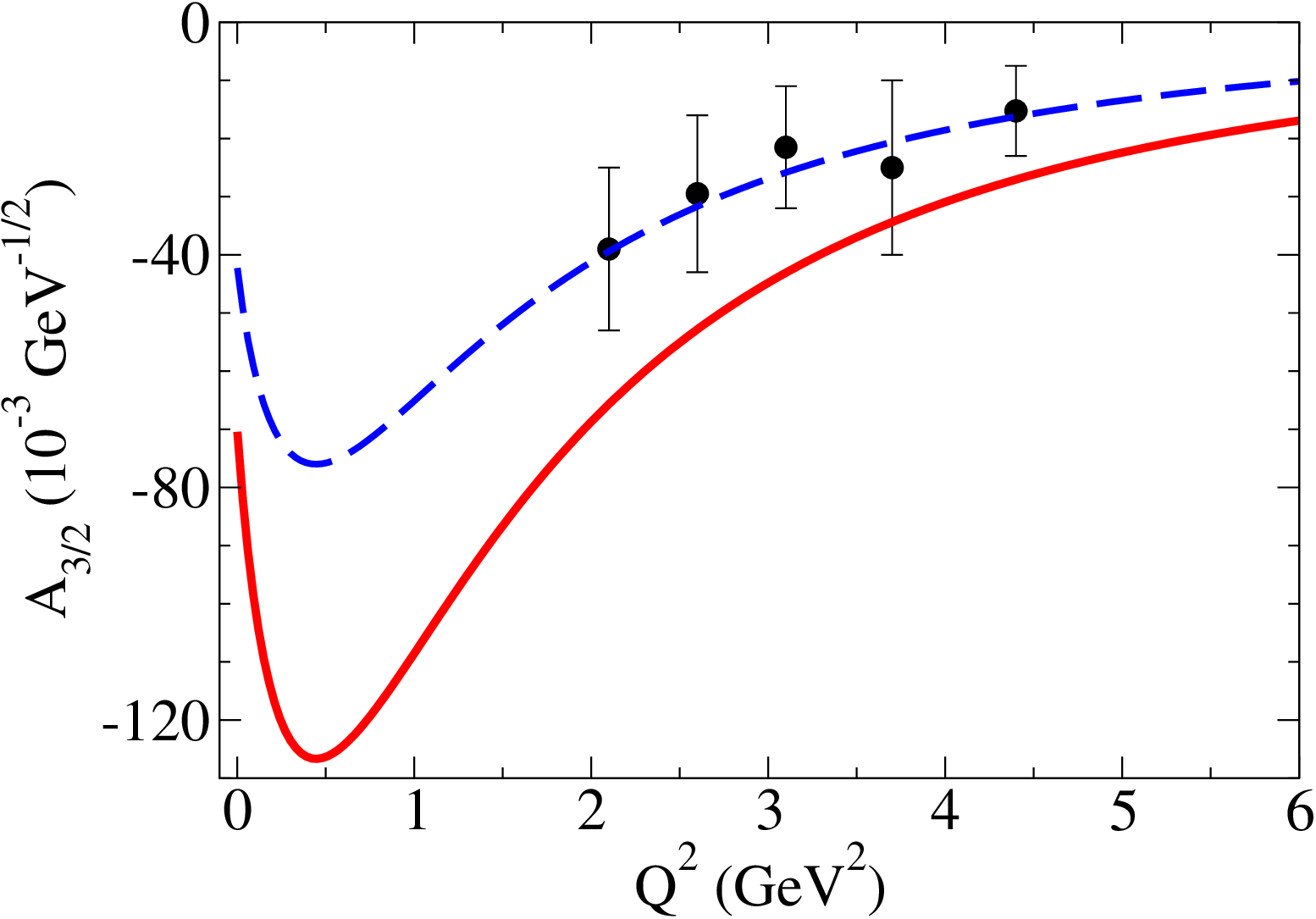} 
\caption{\footnotesize
  Calculations of the $\gamma^\ast N \to \Delta(1600)$ transition amplitudes.
  The data are from JLab/CLAS~\cite{Mokeev23a}. 
\label{fig-Roper-Amps4}}
\end{figure}

The original estimate for the transverse amplitudes, corrected by the sign of the data, is presented in Fig.~\ref{fig-Roper-Amps4} with the solid line.
It is clear that our calculations overestimate the novel data in absolute value.
The first estimate can be improved when we combine the original estimate of the valence quark contribution with the intermediate value for the meson cloud ($G_M^{\rm MC} (0) \simeq -0.65$).
With the new estimate, the combination of valence and meson cloud contributions is obtained when we consider the normalization  $Z \simeq 0.5$ for the sum of both contributions~\cite{NSTAR2024}. 
The combined result is represented in Fig.~\ref{fig-Roper-Amps4} by the dashed line.
From the figure, we can conclude that the new data is well described assuming that $\Delta(1600)$ is the radial description of the data, when the meson cloud contributes about 50\% for the amplitudes near $Q^2=0$.

From the analysis of the data, we can conclude also that the $\gamma^\ast N \to \Delta(1600)$ transition is dominated by the magnetic form factor~\cite{Delta1600,NSTAR2024}.
The JLab/CLAS data~\cite{Mokeev23a} for the  $\gamma^\ast N \to \Delta(1600)$ transition is compatible with $G_E \equiv 0$, within the experimental uncertainties~\cite{NSTAR2024}.

%%%%%     Sec6

\section{Discussion about the nature of the Roper \label{sec-Discussion}}

In the previous sections we discussed the properties of the Roper and of the $\gamma^\ast N \to N(1440)$ transition, under the assumption that valence quarks are the relevant degrees of freedom, and that the state can be interpreted as the first radial excitation of the nucleon.
Under these two assumptions, it was possible to use the covariant spectator quark model to make predictions for the transition amplitudes and transition form factors for the large-$Q^2$ region, based exclusively on the parameters determined by the study of the nucleon.
Similar results can be obtained in the same region using a holographic QCD model in leading order (three-quark structure).
Both calculations are in good agreement with the large-$Q^2$ data.

As discussed extensively in the literature, the consideration of the quark degrees of freedom exclusively is insufficient to describe the observed properties of the Roper.
Calculations based on valence quarks are unable to explain why the physical mass ($M_R \simeq 1.4$ GeV) is much smaller than the mass estimated by models ($M_R=1.7$--1.9 GeV)~\cite{Crede13a,Capstick00,Capstick86b,Burkert19a},
and why the decay width ($\Gamma_R \simeq 350$ MeV) is much larger than the other resonances from the first and second resonance regions (typically $\Gamma_R \simeq 150$ MeV)~\cite{Blin19a,Burkert19a}.

These discrepancies can be understood when we consider that the Roper is a hybrid system: a combination of a three-quark state with some components from baryon-meson states~\cite{Burkert04,JDiaz09a,Suzuki10a,Kamano13a,Nakamura15a,Kamano16a}.
Models based on the baryon-meson states describes well the static properties (mass, decay width and branching ratios) and the transition amplitudes in the low-$Q^2$ region.
Calculations based on chiral effective baryon-meson models~\cite{Hernandez02a,Bauer14a,Gelenava18a,Gegelia16a} and baryon-meson coupled-channel reaction models~\cite{Krehl00a,Ronchen13a,Doring09a,Golli18a}, suggest, in contrast, that the Roper can be described as a dynamically generated resonance without any explicit reference to its quark content.   
The different frameworks predict different weights for the contributions associated with the pure valence quark states and to the contributions from the baryon-meson states associated with the meson cloud effects.
It is worth noticing, however, that a pure molecular-type description of the Roper would lead to much softer transition form factors at large $Q^2$, than observed experimentally~\cite{PPNP2024,Burkert19a}.

The understanding of the nature of the Roper requires then a combination of the two mechanisms described above: baryon-meson states in the low-$Q^2$ region, and valence quark contributions in the large-$Q^2$ region.

The consideration of these two mechanisms can be made using baryon-meson coupled-channel reactions models, combined with an interpretation of the baryon structure based on its underlying quark structure.
An example of a formalism with these properties is the ANL-Osaka model.
The model predicts that the $N(1440)$ state can be generated from a bare resonance with a mass $M_R \simeq 1.7$ GeV, when the dressing of the baryon bare cores by the meson cloud is taken into account~\cite{Suzuki10a,Kamano13a,Nakamura15a}.
Another consequence of the meson cloud dressing is a reduction of the mass of the resonance by about 20\%~\cite{Burkert19a}.
These results show how we can conciliate a large mass when we consider the effects of valence quarks, with a lower mass when we take into account meson cloud effects at the hadronic level.

The analysis based on baryon-meson coupled-channel reaction models identifies the nucleon resonances as poles in the complex plane.
In some cases the bare state (determined in the limit when the meson cloud is removed) generates more than one pole when the meson cloud dressing is taken into account.
In the case of the Roper, there are analyses that suggest that the Roper is a combination of two poles, while other analyzes predict only one pole~\cite{Burkert19a,Suzuki10a}.
For more detailed discussions, the readers can consult Refs.~\cite{Ronchen13a,Cutkosky90a,Arndt06a,Arndt85a,Sarantsev08a}.
In the two-poles case, the measurements of the $\gamma^\ast N \to N(1440)$ amplitudes from Sec.~\ref{secResults} should be compared with the estimates of the combinations of the two states.

The hybrid structure of the Roper has also been revealed in studies of lattice QCD in finite volumes.
Evidence of a state associated with the first radial excitation of the nucleon has been found in lattice QCD simulations for large pion masses~\cite{DRoberts14a}.
The determination of the $N^\ast$ spectrum and the mass associated with the  first radial excitation of the nucleon in lattice QCD simulations has been, however, a complicated task.
The order of the second $N\left( \sfrac{1}{2}^+ \right)$ state and the first $N\left( \sfrac{1}{2}^- \right)$ state in lattice QCD simulations is sensitive to the method used (quenched, unquenched, etc.), to the pion mass and to the lattice volume~\cite{Burkert19a,Mathur05a,Mahbub10b,Lang17a,Sun20a,Wu18b,Owa25a}.

The interpretation of the mass spectra in lattice QCD must be done with care, since the energy levels can include three-quark states represented by local $(qqq)$ local interpolating fields, but also mixtures with $(qqq) (q \bar q)$ and $(qqq) (q \bar q) (q \bar q)$ non local interpolating fields, that can be interpreted as baryon-meson rescattering channels~\cite{Lang17a,Wu18b,Owa25a,Liu16a}.
The connection between the energy levels in finite volume lattices and the infinite volume energy levels can be made using the L\"uscher method combined with Hamiltonian effective field theory~\cite{Wu18b,Liu16a,Liu17a}, expressed in terms of bare baryons and baryon-meson channels in finite lattice.
The interpretation of the contributions for the lattice energy levels is obtained when free parameters of the model are calibrated by the  $\pi N$ experimental phase shift data~\cite{Workman12a,GWU}, and the results of the baryon-meson phase shifts obtained in lattice QCD simulations with finite volumes.

In this picture the Roper appears as a state dominated by a strong rescattering between coupled meson-baryon channels, with predominance to the $\sigma N$ channel, combined with a small contribution of a bare three-quark state associated with the first radial excitation of the bare nucleon, with a mass of about 2 GeV~\cite{Wu18b,Owa25a}.
The strong baryon-meson rescattering helps to describe the $\pi N$ experimental phase shifts and the $\pi N$ lattice QCD phase shifts, while the falloffs of the transition amplitudes at large $Q^2$ can be interpreted as a consequence of the small mixture with a bare three-quark state with a large mass~\cite{Wu18b}.
Our calculations are also compatible with the Roper resonance with a larger effective mass.

From the analysis based on dynamical baryon-meson coupled-channel reaction models combined with the quark structure of bare cores (ANL-Osaka model), and the analysis of the lattice QCD based on the baryon-meson phase shifts in finite volume, one has to conclude that the controversy on the structure of the Roper is not yet fully decided.

Future experiments on higher mass  $N\left(\frac{1}{2}^+\right)$ resonances may help to identify the structure of the other radial excitations of the nucleon.
Meanwhile, the extension of lattice QCD studies to larger volumes and to a basis with more baryon-meson states, may help to conciliate model calculations based on constituent quarks with the structure revealed in finite lattice simulations.
Combined studies based on quark degrees of freedom, dynamical baryon-meson coupled-channel reaction models, and lattice QCD studies with increasing number of bare quark states and baryon-meson channels may help to quantify the weight of the valence quark and the baryon-meson components in the different physical processes, in the low-$Q^2$ and in the large-$Q^2$ regions.

\section{Outlook and conclusions \label{sec-Conclusions}}

The Roper resonance, discovered by David Roper more than 60 years ago, is a special case among the $N^\ast$ resonances.
Recent measurements of the nucleon to Roper transition amplitudes at relatively large $Q^2$ reveal a structure associated with a core of three quarks, while the mass, decay width, and branching ratios suggest a significant role of the baryon-meson states related to the meson cloud dressing of the baryon cores, and a molecular-type nature.

The dominant interpretation of the Roper is that the state is mainly a bound state of three dressed quarks corresponding to the first radial excitation of the nucleon, but that the meson cloud dressing associated with the $\pi N$, $\pi \Delta$ and $\sigma N$ channels also contributes to the physical properties of the resonance.
The meson cloud structure contributes to the large decay width and to the reduction of the physical mass compared with the mass estimated by quark models.
This picture is consistent with calculations of transition amplitudes and form factors with frameworks based on quark degrees of freedom, including Dyson-Schwinger equations, and with calculations based on effective field theory and baryon-meson coupled-channel reaction models.
At the moment, however,  there are still some discrepancies with lattice QCD calculations based on the analysis of the phase shifts on the dominant decay channels.

Future measurements of the $\gamma^\ast N \to N(1440)$ transition amplitudes above $Q^2= 5$ GeV$^2$ may help to conclude if the falloffs observed for the amplitudes are in fact associated with the radial excitation of the nucleon, or just a shadow of the three-quark core of the nucleon hidden on a cloud of mesons.
Furthermore, measurements of the $\gamma^\ast N \to N^\ast$ transition amplitudes, where $N^\ast$ is a nucleon resonance associated with the second radial excitation of the nucleon [tentatively the resonance $N(1880)$], for large $Q^2$, may also help to understand if the radial structure of nucleon resonances $N \left( \frac{1}{2}^+ \right)$ may be confirmed experimentally.
Calculations based on quark degrees of freedom and holographic QCD models for the first and second radial excitations of the nucleon, above $Q^2=4$ GeV$^2$, will be very useful for the interpretation of the structure associated with those states.

Experiments in the low-$Q^2$ region, in the range $Q^2=0.05$--0.25 GeV$^2$, will be fundamental for the effective determination of the shape of the $A_{1/2}$ and $S_{1/2}$ amplitudes, and to conciliate the measurements from JLab/CLAS with the measurements from MAMI.
One can then confirm if the theoretical constraints at the pseudothreshold have a visible impact on the amplitudes near the photon point.

\section*{Acknowledgments}

G.R.~would like to thank Derek Leinweber, 
Hiroyuki Kamano, Toru Sato, Alex Sibirtsev,
Igor Strakovsky and Tony Thomas for the useful discussions. 
This work was supported by the National
Research Foundation of Korea (Grant  No.~RS-2021-NR060129).


\begin{thebibliography}{00}


\bibitem{PDG2024}
S.~Navas \textit{et al.} [Particle Data Group],
``Review of particle physics,''
\href{https://doi.org/10.1103/PhysRevD.110.030001}{\B Phys. Rev. D \textbf{110}, 030001 (2024).}
%%doi:10.1103/PhysRevD.110.030001
%2947 citations counted in INSPIRE as of 22 Oct 2025




\bibitem{Roper64a}
L.~D.~Roper,
``Evidence for a $P_{11}$ Pion-Nucleon Resonance at 556 MeV,''
\href{https://doi.org/10.1103/PhysRevLett.12.340}{\B Phys. Rev. Lett. \textbf{12}, 340 (1964).}
%%doi:10.1103/PhysRevLett.12.340
%228 citations counted in INSPIRE as of 22 Oct 2025

\bibitem{Roper65a}
L.~D.~Roper, R.~M.~Wright and B.~T.~Feld,
``Energy-Dependent Pion-Nucleon Phase-Shift Analysis,''
\href{https://doi.org/10.1103/PhysRev.138.B190}{\B 
Phys. Rev. \textbf{138}, B190 (1965).}
%%doi:10.1103/PhysRev.138.B190
%217 citations counted in INSPIRE as of 22 Oct 2025



\bibitem{PPNP2024}
G.~Ramalho and M.~T.~Pe{\~n}a,
``Electromagnetic transition form factors of baryon resonances,''
\href{https://doi.org/10.1016/j.ppnp.2024.104097}{\B 
Prog. Part. Nucl. Phys. \textbf{136}, 104097 (2024).}
%%doi:10.1016/j.ppnp.2024.104097
%%[arXiv:2306.13900 [hep-ph]].
%35 citations counted in INSPIRE as of 22 Oct 2025


\bibitem{NSTAR}
I.~G.~Aznauryan, A.~Bashir, V.~Braun, S.~J.~Brodsky, V.~D.~Burkert, L.~Chang, C.~Chen, B.~El-Bennich, I.~C.~Cloet and P.~L.~Cole, \textit{et al.}
``Studies of Nucleon Resonance Structure in Exclusive Meson Electroproduction,''
\href{https://doi.org/10.1142/S0218301313300154}{\B 
Int. J. Mod. Phys. E \textbf{22}, 1330015 (2013).}
%%doi:10.1142/S0218301313300154
%%[arXiv:1212.4891 [nucl-th]].
%243 citations counted in INSPIRE as of 22 Oct 2025


\bibitem{Aznauryan12a}
I.~G.~Aznauryan and V.~D.~Burkert,
``Electroexcitation of nucleon resonances,''
\href{https://doi.org/10.1016/j.ppnp.2011.08.001}{\B 
Prog. Part. Nucl. Phys. \textbf{67}, 1 (2012).}
%%doi:10.1016/j.ppnp.2011.08.001
%%[arXiv:1109.1720 [hep-ph]].
%247 citations counted in INSPIRE as of 22 Oct 2025


\bibitem{CLAS09}
I.~G.~Aznauryan \textit{et al.} [CLAS],
``Electroexcitation of nucleon resonances from \mbox{CLAS data on single pion electroproduction,''
%Phys. Rev. C \textbf{80}, 055203 (2009).
\href{https://doi.org/10.1103/PhysRevC.80.055203}{\B Phys. Rev. C \textbf{80}, 055203 (2009).}}
%%doi:10.1103/PhysRevC.80.055203
%%[arXiv:0909.2349 [nucl-ex]].
%305 citations counted in INSPIRE as of 22 Oct 2025


\bibitem{CLAS12}
V.~I.~Mokeev \textit{et al.} [CLAS],
``Experimental Study of the $P_{11}(1440)$ and $D_{13}(1520)$ resonances from CLAS data on $ep \rightarrow e'\pi^{+} \pi^{-} p'$,''
\href{https://doi.org/10.1103/PhysRevC.86.035203}{\B Phys. Rev. C \textbf{86}, 035203 (2012).}
%%doi:10.1103/PhysRevC.86.035203
%%[arXiv:1205.3948 [nucl-ex]].
%147 citations counted in INSPIRE as of 22 Oct 2025


\bibitem{CLAS16a}
V.~I.~Mokeev, V.~D.~Burkert, D.~S.~Carman, L.~Elouadrhiri, G.~V.~Fedotov, E.~N.~Golovatch, R.~W.~Gothe, K.~Hicks, B.~S.~Ishkhanov and E.~L.~Isupov, \textit{et al.}
``New Results from the Studies of the $N(1440)1/2^+$, $N(1520)3/2^-$, and $\Delta(1620)1/2^-$ Resonances in Exclusive $ep \to e'p' \pi^+ \pi^-$ Electroproduction with the CLAS Detector,''
\href{https://doi.org/10.1103/PhysRevC.93.025206}{\B Phys. Rev. C \textbf{93}, 025206 (2016).}
%%doi:10.1103/PhysRevC.93.025206
%%[arXiv:1509.05460 [nucl-ex]].
%107 citations counted in INSPIRE as of 22 Oct 2025





\bibitem{Mokeev23a}
V.~I.~Mokeev, P.~Achenbach, V.~D.~Burkert, D.~S.~Carman, R.~W.~Gothe, A.~N.~Hiller Blin, E.~L.~Isupov, K.~Joo, K.~Neupane and A.~Trivedi,
``First Results on Nucleon Resonance Electroexcitation Amplitudes from $ep\to e'\pi^+\pi^-p'$ Cross Sections at $W$ = 1.4-1.7 GeV and $Q^2$ = 2.0-5.0 GeV$^2$,''
\href{https://doi.org/10.1103/PhysRevC.108.025204}{\B Phys. Rev. C \textbf{108}, 025204 (2023).}
%%doi:10.1103/PhysRevC.108.025204
%%[arXiv:2306.13777 [nucl-ex]].
%24 citations counted in INSPIRE as of 22 Oct 2025




\bibitem{Hernandez02a}
E.~Hernandez, E.~Oset and M.~J.~Vicente Vacas,
``The Two pion decay of the Roper resonance,''
\href{https://doi.org/10.1103/PhysRevC.66.065201}{\B Phys. Rev. C \textbf{66}, 065201 (2002).}
%%doi:10.1103/PhysRevC.66.065201
%%[arXiv:nucl-th/0209009 [nucl-th]].
%32 citations counted in INSPIRE as of 26 Oct 2025












  \bibitem{CELSIUS08a}
T.~Skorodko \textit{et al.} [CELSIUS WASA],
``Excitation of the Roper resonance in single- and double-pion production in nucleon-nucleon collisions,''
\href{https://doi.org/10.1140/epja/i2008-10569-6}{\B Eur. Phys. J. A \textbf{35}, 317 (2008).}
%%doi:10.1140/epja/i2008-10569-6
%64 citations counted in INSPIRE as of 26 Oct 2025


\bibitem{ARuso98a}
L.~Alvarez-Ruso, E.~Oset and E.~Hernandez,
%``Theoretical study of the N N ---{\ensuremath{>}} N N pi pi reaction,''
``Theoretical study of the $NN \to NN \pi \pi$ reaction,''
\href{https://doi.org/10.1016/S0375-9474(98)00126-2}{\B Nucl. Phys. A \textbf{633}, 519 (1998).}
%%doi:10.1016/S0375-9474(98)00126-2
%%[arXiv:nucl-th/9706046 [nucl-th]].
%142 citations counted in INSPIRE as of 26 Oct 2025





\bibitem{Cao10a}
X.~Cao, B.~S.~Zou and H.~S.~Xu,
``Phenomenological analysis of the double pion production in nucleon-nucleon collisions up to 2.2 GeV,''
\href{https://doi.org/10.1103/PhysRevC.81.065201}{\B Phys. Rev. C \textbf{81}, 065201 (2010).}
%%doi:10.1103/PhysRevC.81.065201
%%[arXiv:1004.0140 [nucl-th]].
%86 citations counted in INSPIRE as of 26 Oct 2025



\bibitem{Blin19a}
A.~N.~Hiller Blin, V.~Mokeev, M.~Albaladejo, C.~Fern{\'a}ndez-Ram{\'\i}rez, V.~Mathieu, A.~Pilloni, A.~Szczepaniak, V.~D.~Burkert, V.~V.~Chesnokov and A.~A.~Golubenko, \textit{et al.}
``Nucleon resonance contributions to unpolarised inclusive electron scattering,''
\href{https://doi.org/10.1103/PhysRevC.100.035201}{\B Phys. Rev. C \textbf{100}, 035201 (2019).}
%%doi:10.1103/PhysRevC.100.035201
%%[arXiv:1904.08016 [hep-ph]].
%41 citations counted in INSPIRE as of 22 Oct 2025




\bibitem{Eichmann18}
G.~Eichmann and G.~Ramalho,
``Nucleon resonances in Compton scattering,''
\href{https://doi.org/10.1103/PhysRevD.98.093007}{\B Phys. Rev. D \textbf{98}, 093007 (2018).}
%%doi:10.1103/PhysRevD.98.093007
%%[arXiv:1806.04579 [hep-ph]].
%30 citations counted in INSPIRE as of 22 Oct 2025




\bibitem{Crede13a}
V.~Crede and W.~Roberts,
``Progress towards understanding baryon resonances,''
\href{https://doi.org/10.1088/0034-4885/76/7/076301}{\B Rept. Prog. Phys. \textbf{76}, 076301 (2013).}
%%doi:10.1088/0034-4885/76/7/076301
%%[arXiv:1302.7299 [nucl-ex]].
%294 citations counted in INSPIRE as of 24 Oct 2025



\bibitem{Capstick00}
S.~Capstick and W.~Roberts,
``Quark models of baryon masses and decays,''
\href{https://doi.org/10.1016/S0146-6410(00)00109-5}{\B Prog. Part. Nucl. Phys. \textbf{45}, S241 (2000).}
%%doi:10.1016/S0146-6410(00)00109-5
%%[arXiv:nucl-th/0008028 [nucl-th]].
%498 citations counted in INSPIRE as of 22 Oct 2025



\bibitem{Capstick86b}
S.~Capstick and N.~Isgur,
``Baryons in a relativized quark model with chromodynamics,''
\href{https://doi.org/10.1103/physrevd.34.2809}{\B Phys. Rev. D \textbf{34}, 2809 (1986).}
%%doi:10.1103/physrevd.34.2809
%1546 citations counted in INSPIRE as of 27 Oct 2025
      %%    Spectrum baryons   KARK-ISGUR    Capstick-Isgur



\bibitem{Isgur79b}
N.~Isgur and G.~Karl,
``Positive Parity Excited Baryons in a Quark Model with Hyperfine Interactions,''
\href{https://doi.org/10.1103/PhysRevD.19.2653}{\B Phys. Rev. D \textbf{19}, 2653 (1979)}
[erratum: Phys. Rev. D \textbf{23}, 817 (1981)].
%doi:10.1103/PhysRevD.19.2653
%1108 citations counted in INSPIRE as of 23 Oct 2025
%%%    DOI: https://doi.org/10.1103/PhysRevD.23.817


\bibitem{Burkert19a}
V.~D.~Burkert and C.~D.~Roberts,
``Colloquium : Roper resonance: Toward a solution to the fifty year puzzle,''
\href{https://doi.org/10.1103/RevModPhys.91.011003}{\B Rev. Mod. Phys. \textbf{91}, 011003 (2019).}
%%doi:10.1103/RevModPhys.91.011003
%%[arXiv:1710.02549 [nucl-ex]].
%135 citations counted in INSPIRE as of 22 Oct 2025






\bibitem{Li92a}
Z.~p.~Li, V.~Burkert and Z.~j.~Li,
``Electroproduction of the Roper resonance as a hybrid state,''
\href{https://doi.org/10.1103/PhysRevD.46.70}{\B Phys. Rev. D \textbf{46}, 70 (1992).}
%%doi:10.1103/PhysRevD.46.70
%192 citations counted in INSPIRE as of 26 Oct 2025



\bibitem{Alberto01a}
P.~Alberto, M.~Fiolhais, B.~Golli and J.~Marques,
``N* electroproduction amplitudes in a model with dynamical confinement,''
%Phys. Lett. B \textbf{523}, 273 (2001).
\href{https://doi.org/10.1016/S0370-2693(01)01348-X}{\B Phys. Lett. B \textbf{523}, 273 \\ (2001).}
%%doi:10.1016/S0370-2693(01)01348-X
%%[arXiv:hep-ph/0103171 [hep-ph]].
%25 citations counted in INSPIRE as of 26 Oct 2025
    %%  cluster of three quarks in radial orbital configuration
    %%  surrounded by π- and σ -meson clouds




\bibitem{Kaulfuss85a}
U.~B.~Kaulfuss and U.~G.~Meissner,
``The Breathing Mode of the Modified Skyrmion,''
\href{https://doi.org/10.1016/0370-2693(85)90583-0}{\B Phys. Lett. B \textbf{154}, 193 (1985).}
%%doi:10.1016/0370-2693(85)90583-0
%29 citations counted in INSPIRE as of 26 Oct 2025
    %%  Skyrmion -- breathing mode


\bibitem{Mattis85a}
M.~P.~Mattis and M.~Karliner,
``The Baryon Spectrum of the Skyrme Model,''
\href{https://doi.org/10.1103/PhysRevD.31.2833}{\B Phys. Rev. D \textbf{31}, 2833 (1985).}
%%doi:10.1103/PhysRevD.31.2833
%221 citations counted in INSPIRE as of 26 Oct 2025





\bibitem{Krehl00a}
O.~Krehl, C.~Hanhart, S.~Krewald and J.~Speth,
``What is the structure of the Roper resonance?,''
\href{https://doi.org/10.1103/PhysRevC.62.025207}{\B Phys. Rev. C \textbf{62}, 025207 (2000).}
%%doi:10.1103/PhysRevC.62.025207
%%[arXiv:nucl-th/9911080 [nucl-th]].
%264 citations counted in INSPIRE as of 26 Oct 2025


\bibitem{Ronchen13a}
D.~Ronchen, M.~Doring, F.~Huang, H.~Haberzettl, J.~Haidenbauer, C.~Hanhart, S.~Krewald, U.~G.~Meissner and K.~Nakayama,
``Coupled-channel dynamics in the reactions
$\pi N \to  \pi N$, $\eta N$, $K \Lambda$, $K \Sigma$,''
\href{https://doi.org/10.1140/epja/i2013-13044-5}{\B Eur. Phys. J. A \textbf{49}, 44 (2013).}
%doi:10.1140/epja/i2013-13044-5
%%[arXiv:1211.6998 [nucl-th]].
%218 citations counted in INSPIRE as of 24 Oct 2025



\bibitem{Doring09a}
M.~Doring, C.~Hanhart, F.~Huang, S.~Krewald and U.~G.~Meissner,
``Analytic properties of the scattering amplitude and resonances parameters in a meson exchange model,''
\href{https://doi.org/10.1016/j.nuclphysa.2009.08.010}{\B Nucl. Phys. A \textbf{829}, 170 (2009).}
%%doi:10.1016/j.nuclphysa.2009.08.010
%%[arXiv:0903.4337 [nucl-th]].
%185 citations counted in INSPIRE as of 26 Oct 2025


\bibitem{Wang24}
Y.~F.~Wang \textit{et al.} [J{\"u}lich-Bonn-Washington],
``Global Data-Driven Determination of Baryon Transition Form Factors,''
%Phys. Rev. Lett. \textbf{133}, 101901 (2024).
\href{https://doi.org/10.1103/PhysRevLett.133.101901}{\B Phys. Rev. Lett. \textbf{133}, 101901 \\(2024).}
%%doi:10.1103/PhysRevLett.133.101901
%%[arXiv:2404.17444 [nucl-th]].
%8 citations counted in INSPIRE as of 23 Oct 2025



\bibitem{Doring25a}
M.~D{\"o}ring, J.~Haidenbauer, M.~Mai and T.~Sato,
``Dynamical coupled-channel models for hadron dynamics,''
\href{https://doi.org/10.1016/j.ppnp.2025.104213}{\B Prog. Part. Nucl. Phys. \textbf{146}, 104213 (2026).}
%%doi:10.1016/j.ppnp.2025.104213
%%[arXiv:2505.02745 [nucl-th]].
%19 citations counted in INSPIRE as of 18 Nov 2025
  

\bibitem{Golli18a}
B.~Golli, H.~Osmanovi{\'c}, S.~{\v{S}}irca and A.~{\v{S}}varc,
``Genuine quark state versus dynamically generated structure for the Roper resonance,''
\href{https://doi.org/10.1103/PhysRevC.97.035204}{\B Phys. Rev. C \textbf{97},  035204 (2018).}
%%doi:10.1103/PhysRevC.97.035204
%%[arXiv:1709.09025 [hep-ph]].
%16 citations counted in INSPIRE as of 24 Oct 2025







\bibitem{CLAS08}
I.~G.~Aznauryan \textit{et al.} [CLAS],
``Electroexcitation of the Roper resonance for 1.7 $< Q^2 < $ 4.5 GeV$^2$
in $\vec{e} p \to  e n \pi^+$,''
\href{https://doi.org/10.1103/PhysRevC.78.045209}{\B Phys. Rev. C \textbf{78}, 045209 (2008).}
%%doi:10.1103/PhysRevC.78.045209
%%[arXiv:0804.0447 [nucl-ex]].
%105 citations counted in INSPIRE as of 23 Oct 2025



\bibitem{Nucleon}
F.~Gross, G.~Ramalho and M.~T.~Pe\~na,
``A Pure S-wave covariant model for the nucleon,''
\href{https://doi.org/10.1103/PhysRevC.77.015202}{\B Phys. Rev. C \textbf{77}, 015202 (2008).}
%%doi:10.1103/PhysRevC.77.015202
%%[arXiv:nucl-th/0606029 [nucl-th]].
%114 citations counted in INSPIRE as of 22 Oct 2025



\bibitem{NDelta}
G.~Ramalho, M.~T.~Pe\~na and F.~Gross,
``A Covariant model for the nucleon and the $\Delta$,''
\href{https://doi.org/10.1140/epja/i2008-10599-0}{\B Eur. Phys. J. A \textbf{36}, 329 (2008).}
%%doi:10.1140/epja/i2008-10599-0
%%[arXiv:0803.3034 [hep-ph]].
%72 citations counted in INSPIRE as of 22 Oct 2025


\bibitem{NDeltaD}
G.~Ramalho, M.~T.~Pe\~na and F.~Gross,
``D-state effects in the electromagnetic $N \Delta$ transition,''
\href{https://doi.org/10.1103/PhysRevD.78.114017}{\B Phys. Rev. D \textbf{78}, 114017 (2008).}
%%doi:10.1103/PhysRevD.78.114017
%%[arXiv:0810.4126 [hep-ph]].
%75 citations counted in INSPIRE as of 22 Oct 2025



\bibitem{Lattice}
G.~Ramalho and M.~T.~Pe\~na,
``Nucleon and  $\gamma N \to \Delta$ lattice form factors in a constituent quark model,''
\href{https://doi.org/10.1088/0954-3899/36/11/115011}{\B J. Phys. G \textbf{36}, 115011 (2009).}
%%doi:10.1088/0954-3899/36/11/115011
%%[arXiv:0812.0187 [hep-ph]].
%54 citations counted in INSPIRE as of 22 Oct 2025


\bibitem{LatticeD}
G.~Ramalho and M.~T.~Pe\~na,
``Valence quark contribution for the $\gamma N \to \Delta$ 
quadrupole transition extracted from lattice QCD,''
\href{https://doi.org/10.1103/PhysRevD.80.013008}{\B Phys. Rev. D \textbf{80}, 013008 (2009).}
%%doi:10.1103/PhysRevD.80.013008
%%[arXiv:0901.4310 [hep-ph]].
%68 citations counted in INSPIRE as of 22 Oct 2025



\bibitem{Timelike}
G.~Ramalho and M.~T.~Pe\~na,
``Timelike  $\gamma^\ast N \to \Delta$ form factors and Delta Dalitz decay,''
\href{https://doi.org/10.1103/PhysRevD.85.113014}{\B Phys. Rev. D \textbf{85}, 113014 (2012).}
%%doi:10.1103/PhysRevD.85.113014
%%[arXiv:1205.2575 [hep-ph]].
%61 citations counted in INSPIRE as of 22 Oct 2025


\bibitem{Timelike2}
G.~Ramalho, M.~T.~Pe\~na, J.~Weil, H.~van Hees and U.~Mosel,
``Role of the pion electromagnetic form factor in the $\Delta(1232) \to \gamma^\ast N$ timelike transition,''
\href{https://doi.org/10.1103/PhysRevD.93.033004}{\B Phys. Rev. D \textbf{93}, 033004 (2016).}
%%doi:10.1103/PhysRevD.93.033004
%%[arXiv:1512.03764 [hep-ph]].
%47 citations counted in INSPIRE as of 22 Oct 2025



\bibitem{Siegert1}
G.~Ramalho,
``Parametrizations of the $\gamma^\ast N \to \Delta(1232)$ quadrupole form factors and Siegert{\textquoteright}s theorem,''
\href{https://doi.org/10.1103/PhysRevD.94.114001}{\B Phys. Rev. D \textbf{94}, 114001 (2016).}
%%doi:10.1103/PhysRevD.94.114001
%%[arXiv:1606.03042 [hep-ph]].
%25 citations counted in INSPIRE as of 22 Oct 2025


\bibitem{Siegert3}
G.~Ramalho,
``New low-$Q^2$ measurements of the $\gamma^\ast N \to \Delta(1232)$ Coulomb quadrupole form factor, pion cloud parametrizations and Siegert's theorem,''
\href{https://doi.org/10.1140/epja/i2018-12514-6}{\B Eur. Phys. J. A \textbf{54}, 75 (2018).}
%%doi:10.1140/epja/i2018-12514-6
%%[arXiv:1709.07412 [hep-ph]].
%20 citations counted in INSPIRE as of 22 Oct 2025



\bibitem{Delta-shape}
G.~Ramalho, M.~T.~Pe\~na and A.~Stadler,
``The shape of the $\Delta$ baryon in a covariant spectator quark model,''
\href{https://doi.org/10.1103/PhysRevD.86.093022}{\B Phys. Rev. D \textbf{86}, 093022 (2012).}
%%doi:10.1103/PhysRevD.86.093022
%%[arXiv:1207.4392 [nucl-th]].
%26 citations counted in INSPIRE as of 24 Oct 2025




\bibitem{Omega}
G.~Ramalho, K.~Tsushima and F.~Gross,
``A Relativistic quark model for the $\Omega^-$ electromagnetic form factors,''
\href{https://doi.org/10.1103/PhysRevD.80.033004}{\B Phys. Rev. D \textbf{80}, 033004 (2009).}
%%doi:10.1103/PhysRevD.80.033004
%%[arXiv:0907.1060 [hep-ph]].
%76 citations counted in INSPIRE as of 22 Oct 2025


\bibitem{NSTAR2017}
G.~Ramalho,
``$N^\ast$ Form Factors based on a Covariant Quark Model,''
\href{https://doi.org/10.1007/s00601-018-1412-9}{\B Few Body Syst. \textbf{59}, 92 (2018).}
%%doi:10.1007/s00601-018-1412-9
%%[arXiv:1801.01476 [hep-ph]].
%22 citations counted in INSPIRE as of 22 Oct 2025




\bibitem{Nucleon2}
F.~Gross, G.~Ramalho and M.~T.~Pe\~na,
``Covariant nucleon wave function with S, D, and P-state components,''
\href{https://doi.org/10.1103/PhysRevD.85.093005}{\B Phys. Rev. D \textbf{85}, 093005 (2012).}
%%doi:10.1103/PhysRevD.85.093005
%%[arXiv:1201.6336 [hep-ph]].
%42 citations counted in INSPIRE as of 22 Oct 2025


\bibitem{Aznauryan07}
I.~G.~Aznauryan,
``Electroexcitation of the Roper resonance in the relativistic quark models,''
\href{https://doi.org/10.1103/PhysRevC.76.025212}{\B Phys. Rev. C \textbf{76}, 025212 (2007).}
%%doi:10.1103/PhysRevC.76.025212
%%[arXiv:nucl-th/0701012 [nucl-th]].
%82 citations counted in INSPIRE as of 22 Oct 2025



\bibitem{JDiaz04a}
B.~Julia-Diaz, D.~O.~Riska and F.~Coester,
``Baryon form-factors of relativistic constituent quark models,''
\href{https://doi.org/10.1103/PhysRevC.69.035212}{\B Phys. Rev. C \textbf{69}, 035212 (2004)}
[erratum: Phys. Rev. C \textbf{75}, 069902 (2007)].
%%doi:10.1103/PhysRevC.69.035212
%%[arXiv:hep-ph/0312169 [hep-ph]].
%82 citations counted in INSPIRE as of 23 Oct 2025




\bibitem{N1440}
G.~Ramalho and K.~Tsushima,
``Valence quark contributions for the
$\gamma N \to P_{11}(1440)$ form factors,''
\href{https://doi.org/10.1103/PhysRevD.81.074020}{\B 
Phys. Rev. D \textbf{81}, 074020 (2010).}
%%doi:10.1103/PhysRevD.81.074020
%%[arXiv:1002.3386 [hep-ph]].
%65 citations counted in INSPIRE as of 22 Oct 2025










\bibitem{Burkert04}
V.~D.~Burkert and T.~S.~H.~Lee,
``Electromagnetic meson production in the nucleon resonance region,''
\href{https://doi.org/10.1142/S0218301304002545}{\B Int. J. Mod. Phys. E \textbf{13}, 1035 (2004).}
%%doi:10.1142/S0218301304002545
%%[arXiv:nucl-ex/0407020 [nucl-ex]].
%217 citations counted in INSPIRE as of 22 Oct 2025


\bibitem{Tiator04}
L.~Tiator, D.~Drechsel, S.~Kamalov, M.~M.~Giannini, E.~Santopinto and A.~Vassallo,
``Electroproduction of nucleon resonances,''
\href{https://doi.org/10.1140/epjad/s2004-03-009-9}{\B 
Eur. Phys. J. A \textbf{19}, 55 (2004).}
%%doi:10.1140/epjad/s2004-03-009-9
%%[arXiv:nucl-th/0310041 [nucl-th]].
%114 citations counted in INSPIRE as of 22 Oct 2025


\bibitem{Drechsel07}
D.~Drechsel, S.~S.~Kamalov and L.~Tiator,
``Unitary Isobar Model - MAID2007,''
\href{https://doi.org/10.1140/epja/i2007-10490-6}{\B Eur. Phys. J. A \textbf{34}, 69 (2007).}
%%doi:10.1140/epja/i2007-10490-6
%%[arXiv:0710.0306 [nucl-th]].
%582 citations counted in INSPIRE as of 22 Oct 2025


\bibitem{N1710}
G.~Ramalho and K.~Tsushima,
``$\gamma^\ast N \to N(1710)$ transition at high momentum transfer,''
\href{https://doi.org/10.1103/PhysRevD.89.073010}{\B Phys. Rev. D \textbf{89}, 073010 (2014).}
%%doi:10.1103/PhysRevD.89.073010
%%[arXiv:1402.3234 [hep-ph]].
%25 citations counted in INSPIRE as of 22 Oct 2025




\bibitem{Delta1600}
G.~Ramalho and K.~Tsushima,
``A Model for the $\Delta(1600)$ resonance and $\gamma N \to \Delta(1600)$ transition,''
\href{https://doi.org/10.1103/PhysRevD.82.073007}{\B Phys. Rev. D \textbf{82}, 073007 (2010).}
%%doi:10.1103/PhysRevD.82.073007
%%[arXiv:1008.3822 [hep-ph]].
%50 citations counted in INSPIRE as of 22 Oct 2025






\bibitem{Roper-AdS1}
G.~Ramalho and D.~Melnikov,
``Valence quark contributions for the $\gamma^\ast N \to N(1440)$ form factors from light-front holography,''
\href{https://doi.org/10.1103/PhysRevD.97.034037}{\B Phys. Rev. D \textbf{97}, 034037 (2018).}
%%doi:10.1103/PhysRevD.97.034037
%%[arXiv:1703.03819 [hep-ph]].
%26 citations counted in INSPIRE as of 22 Oct 2025




\bibitem{Roper-AdS2}
G.~Ramalho,
``Analytic parametrizations of the $\gamma^\ast N \to N(1440)$ form factors inspired by light-front holography,''
\href{https://doi.org/10.1103/PhysRevD.96.054021}{\B Phys. Rev. D \textbf{96}, 054021 (2017).}
%doi:10.1103/PhysRevD.96.054021
%%[arXiv:1706.05707 [hep-ph]].
%18 citations counted in INSPIRE as of 22 Oct 2025





\bibitem{Devenish76}
R.~C.~E.~Devenish, T.~S.~Eisenschitz and J.~G.~Korner,
``Electromagnetic $N-N^\ast$ transition form factors,''
\href{https://doi.org/10.1103/PhysRevD.14.3063}{\B Phys. Rev. D \textbf{14}, 3063 (1976).}
%%doi:10.1103/PhysRevD.14.3063
%142 citations counted in INSPIRE as of 23 Oct 2025



\bibitem{Tiator16}
L.~Tiator,
``Pion Electroproduction and Siegert's Theorem,'' \\
\href{https://doi.org/10.1007/s00601-016-1158-1}{\B Few Body Syst. \textbf{57}, no.11, 1087 (2016).}
%%doi:10.1007/s00601-016-1158-1
%15 citations counted in INSPIRE as of 22 Oct 2025




\bibitem{LowQ2param}
G.~Ramalho,
``Low-$Q^2$ empirical parametrizations of the $N^\ast$ helicity amplitudes,''
\href{https://doi.org/10.1103/PhysRevD.100.114014}{\B Phys. Rev. D \textbf{100}, 114014 (2019).}
%%doi:10.1103/PhysRevD.100.114014
%%[arXiv:1909.00013 [hep-ph]].
%14 citations counted in INSPIRE as of 22 Oct 2025






\bibitem{Gross69a}
F.~Gross,
``Three-dimensional covariant integral equations for low-energy systems,''
\href{https://doi.org/10.1103/PhysRev.186.1448}{\B Phys. Rev. \textbf{186}, 1448 (1969).}
%%doi:10.1103/PhysRev.186.1448
%464 citations counted in INSPIRE as of 23 Oct 2025



\bibitem{Gross99}
F.~Gross,
``Relativistic quantum mechanics and field theory,''
%17 citations counted in INSPIRE as of 23 Oct 2025
Wiley, New York, 1999.
%%   doi:10.1002/9783527617333.
%%  ISBN 978-0-47135-386-7.
%   https://onlinelibrary.wiley.com/doi/book/10.1002/9783527617333
%   DOI:10.1002/9783527617333




\bibitem{Octet}
G.~Ramalho and K.~Tsushima,
``Octet baryon electromagnetic form factors in a relativistic quark model,''
\href{https://doi.org/10.1103/PhysRevD.84.054014}{\B Phys. Rev. D \textbf{84}, 054014 (2011).}
%%doi:10.1103/PhysRevD.84.054014
%%[arXiv:1107.1791 [hep-ph]].
%49 citations counted in INSPIRE as of 22 Oct 2025




\bibitem{OctetDecuplet}
G.~Ramalho and K.~Tsushima,
``Octet to decuplet electromagnetic transition in a relativistic quark model,''
\href{https://doi.org/10.1103/PhysRevD.87.093011}{\B Phys. Rev. D \textbf{87}, 093011 (2013).}
%%doi:10.1103/PhysRevD.87.093011
%%[arXiv:1302.6889 [hep-ph]].
%42 citations counted in INSPIRE as of 22 Oct 2025




\bibitem{OctetDecuplet2}
G.~Ramalho and K.~Tsushima,
``Covariant spectator quark model description of the $\gamma^\ast \Lambda \to \Sigma^0$ transition,''
\href{https://doi.org/10.1103/PhysRevD.86.114030}{\B Phys. Rev. D \textbf{86}, 114030 (2012).}
%%doi:10.1103/PhysRevD.86.114030
%%[arXiv:1210.7465 [hep-ph]].
%30 citations counted in INSPIRE as of 22 Oct 2025


\bibitem{Omega2}
G.~Ramalho and M.~T.~Pe\~na,
``Extracting the $\Omega^-$ electric quadrupole moment from lattice QCD data,''
\href{https://doi.org/10.1103/PhysRevD.83.054011}{\B Phys. Rev. D \textbf{83}, 054011 (2011).}
%%doi:10.1103/PhysRevD.83.054011
%%[arXiv:1012.2168 [hep-ph]].
%34 citations counted in INSPIRE as of 22 Oct 2025



\bibitem{HyperonFF2}
G.~Ramalho, M.~T.~Pe{\~n}a, K.~Tsushima and M.~K.~Cheoun,
``Electromagnetic $|G_E/G_M|$ ratios of hyperons at large timelike $q^2$,''
\href{https://doi.org/10.1016/j.physletb.2024.139060}{\B Phys. Lett. B \textbf{858}, 139060 (2024).}
%%doi:10.1016/j.physletb.2024.139060
%%[arXiv:2407.21397 [hep-ph]].
%10 citations counted in INSPIRE as of 22 Oct 2025


\bibitem{N1520TL}
G.~Ramalho and M.~T.~Pe{\~n}a,
``$\gamma^\ast N \to N(1520)$ form factors in the timelike regime,''
\href{https://doi.org/10.1103/PhysRevD.95.014003}{\B Phys. Rev. D \textbf{95}, 014003 (2017).}
%%doi:10.1103/PhysRevD.95.014003
%%[arXiv:1610.08788 [nucl-th]].
%36 citations counted in INSPIRE as of 22 Oct 2025



\bibitem{N1535TL}
G.~Ramalho and M.~T.~Pe{\~n}a,
``Covariant model for the Dalitz decay of the $N(1535)$ resonance,''
\href{https://doi.org/10.1103/PhysRevD.101.114008}{\B Phys. Rev. D \textbf{101}, 114008 (2020).}
%%doi:10.1103/PhysRevD.101.114008
%%[arXiv:2003.04850 [hep-ph]].
%21 citations counted in INSPIRE as of 22 Oct 2025






\bibitem{Symmetry2025}
G.~Ramalho, K.~Tsushima and M.~K.~Cheoun,
``Electroweak Form Factors of Baryons in Dense Nuclear Matter,''
\href{https://doi.org/10.3390/sym17050681}{\B Symmetry \textbf{17}, no.5,  681 (2025).}
%%doi:10.3390/sym17050681
%%[arXiv:2504.15660 [nucl-th]].
%2 citations counted in INSPIRE as of 22 Oct 2025



\bibitem{GA-medium}
G.~Ramalho, K.~Tsushima and M.~K.~Cheoun,
``Weak interaction axial form factors of the octet baryons in nuclear medium,''
\href{https://doi.org/10.1103/PhysRevD.111.013002}{\B 
Phys. Rev. D \textbf{111}, 013002 (2025).}
%%doi:10.1103/PhysRevD.111.013002
%%[arXiv:2406.07958 [hep-ph]].
%6 citations counted in INSPIRE as of 23 Oct 2025



\bibitem{GA-Octet}
G.~Ramalho and K.~Tsushima,
``Axial form factors of the octet baryons in a covariant quark model,''
\href{https://doi.org/10.1103/PhysRevD.94.014001}{\B Phys. Rev. D \textbf{94}, 014001 (2016).}
%%doi:10.1103/PhysRevD.94.014001
%%[arXiv:1512.01167 [hep-ph]].
%28 citations counted in INSPIRE as of 22 Oct 2025




\bibitem{Octet3}
G.~Ramalho, K.~Tsushima and A.~W.~Thomas,
``Octet Baryon Electromagnetic form Factors in Nuclear Medium,''
\href{https://doi.org/10.1088/0954-3899/40/1/015102}{\B J. Phys. G \textbf{40}, 015102 (2013).}
%%doi:10.1088/0954-3899/40/1/015102
%%[arXiv:1206.2207 [hep-ph]].
%62 citations counted in INSPIRE as of 22 Oct 2025



\bibitem{Hyperons-Dalitz}
G.~Ramalho and K.~Tsushima,
``Meson cloud contributions to the Dalitz decays of decuplet to octet baryons,''
\href{https://doi.org/10.1103/PhysRevD.108.074019}{\B Phys. Rev. D \textbf{108}, 074019 (2023).}
%%doi:10.1103/PhysRevD.108.074019
%%[arXiv:2308.04773 [hep-ph]].
%7 citations counted in INSPIRE as of 22 Oct 2025


\bibitem{SemiR}
G.~Ramalho,
``Semirelativistic approximation to the $\gamma^\ast N \to N(1520)$ and $\gamma^\ast N \to N(1535)$ transition form factors,''
\href{https://doi.org/10.1103/PhysRevD.95.054008}{\B Phys. Rev. D \textbf{95}, 054008 (2017).}
%%doi:10.1103/PhysRevD.95.054008
%%[arXiv:1612.09555 [hep-ph]].
%27 citations counted in INSPIRE as of 22 Oct 2025



\bibitem{Brodsky15a}
S.~J.~Brodsky, G.~F.~de Teramond, H.~G.~Dosch and J.~Erlich,
``Light-Front Holographic QCD and Emerging Confinement,''
\href{https://doi.org/10.1016/j.physrep.2015.05.001}{\B Phys. Rept. \textbf{584}, 1 (2015).}
%%doi:10.1016/j.physrep.2015.05.001
%%[arXiv:1407.8131 [hep-ph]].
%542 citations counted in INSPIRE as of 23 Oct 2025



\bibitem{Maldacena98}
J.~M.~Maldacena,
`The Large $N$ limit of superconformal field theories and supergravity,''
\href{https://doi.org/10.4310/ATMP.1998.v2.n2.a1}{\B Adv. Theor. Math. Phys. \textbf{2}, 231 (1998).}
%%doi:10.4310/ATMP.1998.v2.n2.a1
%%[arXiv:hep-th/9711200 [hep-th]].
%21373 citations counted in INSPIRE as of 24 Oct 2025


\bibitem{Witten98}
E.~Witten,
``Anti de Sitter space and holography,''\\
\href{https://doi.org/10.4310/ATMP.1998.v2.n2.a2}{\B Adv. Theor. Math. Phys. \textbf{2}, 253 (1998).}
%%doi:10.4310/ATMP.1998.v2.n2.a2
%%[arXiv:hep-th/9802150 [hep-th]].
%13616 citations counted in INSPIRE as of 24 Oct 2025



\bibitem{Teramond12a}
G.~F.~de Teramond and S.~J.~Brodsky,
``Excited Baryons in Holographic QCD,''
\href{https://doi.org/10.1063/1.3701207}{\B AIP Conf. Proc. \textbf{1432}, no.1, 168 (2012).}
%%doi:10.1063/1.3701207
%%[arXiv:1108.0965 [hep-ph]].
%48 citations counted in INSPIRE as of 23 Oct 2025

\bibitem{Teramond12b}
G.~F.~de Teramond and S.~J.~Brodsky,
``Hadronic Form Factor Models and Spectroscopy Within the Gauge/Gravity Correspondence,''
\href{https://doi.org/10.48550/arXiv.1203.4025}{\B [arXiv:1203.4025 [hep-ph]].}
%125 citations counted in INSPIRE as of 23 Oct 2025
%% Ferrara International School Niccol\`o Cabeo 2011}: Hadronic Physics",
%   https://doi.org/10.48550/arXiv.1203.4025

\bibitem{Gutsche13a}
T.~Gutsche, V.~E.~Lyubovitskij, I.~Schmidt and A.~Vega,
``Nucleon resonances in AdS/QCD,''
\href{https://doi.org/10.1103/PhysRevD.87.016017}{\B Phys. Rev. D \textbf{87}, 016017 (2013).}
%%doi:10.1103/PhysRevD.87.016017
%%[arXiv:1212.6252 [hep-ph]].
%68 citations counted in INSPIRE as of 23 Oct 2025
%





\bibitem{Stajner17}
S.~{\v{S}}tajner, P.~Achenbach, T.~Beranek, J.~Beri{\v{c}}i{\v{c}}, J.~C.~Bernauer, D.~Bosnar, R.~B{\"o}hm, L.~Correa, A.~Denig and M.~O.~Distler, \textit{et al.}
``Beam-Recoil Polarization Measurement of $\pi^0$ Electroproduction on the Proton in the Region of the Roper Resonance,''
\href{https://doi.org/10.1103/PhysRevLett.119.022001}{\B Phys. Rev. Lett. \textbf{119}, 022001 (2017).}
%%doi:10.1103/PhysRevLett.119.022001
%23 citations counted in INSPIRE as of 24 Oct 2025




\bibitem{Bauer14a}
T.~Bauer, S.~Scherer and L.~Tiator,
``Electromagnetic transition form factors of the Roper resonance in effective field theory,''
\href{https://doi.org/10.1103/PhysRevC.90.015201}{\B Phys. Rev. C \textbf{90}, 015201 (2014).}
%%doi:10.1103/PhysRevC.90.015201
%%[arXiv:1402.0741 [nucl-th]].
%29 citations counted in INSPIRE as of 26 Oct 2025
    %%  calculate EMFF


\bibitem{Gelenava18a}
M.~Gelenava,
``Electromagnetic transition form factors of the Roper resonance in baryon chiral perturbation theory,''
\href{https://doi.org/10.1140/epja/i2018-12523-5}{\B Eur. Phys. J. A \textbf{54}, no.5, 88 (2018).}
%%doi:10.1140/epja/i2018-12523-5
%%[arXiv:1711.03494 [nucl-th]].
%5 citations counted in INSPIRE as of 26 Oct 2025
    %%   calculate EMFF



\bibitem{Golli09a}
B.~Golli, S.~Sirca and M.~Fiolhais,
``Pion electro-production in the Roper region in chiral quark models,''
\href{https://doi.org/10.1140/epja/i2009-10878-2}{\B Eur. Phys. J. A \textbf{42}, 185 (2009).}
%%doi:10.1140/epja/i2009-10878-2
%%[arXiv:0906.2066 [nucl-th]].
%32 citations counted in INSPIRE as of 26 Oct 2025





\bibitem{JDiaz09a}
B.~Julia-Diaz, H.~Kamano, T.~S.~H.~Lee, A.~Matsuyama, T.~Sato and N.~Suzuki,
``Dynamical coupled-channels analysis of $p(e,e' \pi)N$ reactions,''
%%``Dynamical coupled-channels analysis of p(e,e-prime pi)N reactions,''
\href{https://doi.org/10.1103/PhysRevC.80.025207}{\B Phys. Rev. C \textbf{80}, 025207 (2009).}
%%doi:10.1103/PhysRevC.80.025207
%%[arXiv:0904.1918 [nucl-th]].
%76 citations counted in INSPIRE as of 27 Oct 2025
%%  J. Diaz 2009
       %%   EBAC



\bibitem{Suzuki10a}
N.~Suzuki, B.~Julia-Diaz, H.~Kamano, T.~S.~H.~Lee, A.~Matsuyama and T.~Sato,
`Disentangling the Dynamical Origin of $P_{11}$ Nucleon Resonances,''
\href{https://doi.org/10.1103/PhysRevLett.104.042302}{\B Phys. Rev. Lett. \textbf{104}, 042302 (2010).}
%%doi:10.1103/PhysRevLett.104.042302
%%[arXiv:0909.1356 [nucl-th]].
%180 citations counted in INSPIRE as of 24 Oct 2025




\bibitem{Matsuyama06a}
A.~Matsuyama, T.~Sato and T.~S.~H.~Lee,
``Dynamical coupled-channel model of meson production reactions in the nucleon resonance region,''
\href{https://doi.org/10.1016/j.physrep.2006.12.003}{\B Phys. Rept. \textbf{439}, 193 (2007).}
%%doi:10.1016/j.physrep.2006.12.003
%%[arXiv:nucl-th/0608051 [nucl-th]].
%261 citations counted in INSPIRE as of 18 Nov 2025


\bibitem{Kamano13a}
H.~Kamano, S.~X.~Nakamura, T.~S.~H.~Lee and T.~Sato,
``Nucleon resonances within a dynamical coupled-channels model of $\pi N$ and $\gamma N$ reactions,''
\href{https://doi.org/10.1103/PhysRevC.88.035209}{\B 
Phys. Rev. C \textbf{88}, 035209 (2013).}
%%doi:10.1103/PhysRevC.88.035209
%%[arXiv:1305.4351 [nucl-th]].
%260 citations counted in INSPIRE as of 23 Oct 2025



\bibitem{Nakamura15a}
S.~X.~Nakamura, H.~Kamano and T.~Sato,
``Dynamical coupled-channels model for neutrino-induced meson productions in resonance region,''
\href{https://doi.org/10.1103/PhysRevD.92.074024}{\B Phys. Rev. D \textbf{92}, 074024 (2015).}
%%doi:10.1103/PhysRevD.92.074024
%%[arXiv:1506.03403 [hep-ph]].
%103 citations counted in INSPIRE as of 23 Oct 2025



\bibitem{Kamano16a}
H.~Kamano, S.~X.~Nakamura, T.~S.~H.~Lee and T.~Sato,
``Isospin decomposition of $\gamma N \to N^*$ transitions within a dynamical coupled-channels model,''
\href{https://doi.org/10.1103/PhysRevC.94.015201}{\B Phys. Rev. C \textbf{94}, 015201 (2016).}
%%doi:10.1103/PhysRevC.94.015201
%%[arXiv:1605.00363 [nucl-th]].
%66 citations counted in INSPIRE as of 23 Oct 2025





\bibitem{Aznauryan12b}
I.~G.~Aznauryan and V.~D.~Burkert,
``Nucleon electromagnetic form factors and electroexcitation of low lying nucleon resonances in a light-front relativistic quark model,''
\href{https://doi.org/10.1103/PhysRevC.85.055202}{\B Phys. Rev. C \textbf{85}, 055202 (2012).}
%%doi:10.1103/PhysRevC.85.055202
%%[arXiv:1201.5759 [hep-ph]].
%81 citations counted in INSPIRE as of 22 Oct 2025


\bibitem{Obukhovsky14a}
I.~T.~Obukhovsky, A.~Faessler, T.~Gutsche and V.~E.~Lyubovitskij,
``Electromagnetic structure of the nucleon and the Roper resonance in a light-front quark approach,''
\href{https://doi.org/10.1103/PhysRevD.89.014032}{\B Phys. Rev. D \textbf{89}, 014032 (2014).}
%%doi:10.1103/PhysRevD.89.014032
%%[arXiv:1306.3864 [hep-ph]].
%19 citations counted in INSPIRE as of 23 Oct 2025
%%        NOT HOLOGRAPHY



\bibitem{Segovia15a}
J.~Segovia, B.~El-Bennich, E.~Rojas, I.~C.~Cloet, C.~D.~Roberts, S.~S.~Xu and H.~S.~Zong,
``Completing the picture of the Roper resonance,''
\href{https://doi.org/10.1103/PhysRevLett.115.171801}{\B 
Phys. Rev. Lett. \textbf{115}, 171801 (2015).}
%%doi:10.1103/PhysRevLett.115.171801
%%[arXiv:1504.04386 [nucl-th]].
%132 citations counted in INSPIRE as of 22 Oct 2025


\bibitem{Segovia16a}
J.~Segovia and C.~D.~Roberts,
``Dissecting nucleon transition electromagnetic form factors,''
\href{https://doi.org/10.1103/PhysRevC.94.042201}{\B Phys. Rev. C \textbf{94}, 042201 (2016).}
%%doi:10.1103/PhysRevC.94.042201
%%[arXiv:1607.04405 [nucl-th]].
%42 citations counted in INSPIRE as of 22 Oct 2025





\bibitem{JLab-website}
  E.~Isupov,
  ``Fits of the resonances electrocouplings (Jefferson Lab),''\\
  \href{https://userweb.jlab.org/~isupov/couplings/}{\B https://userweb.jlab.org/~isupov/couplings/.}
  %\url{https://userweb.jlab.org/~isupov/couplings/}



  
\bibitem{MAID2011}
L.~Tiator, D.~Drechsel, S.~S.~Kamalov and M.~Vanderhaeghen,
``Electromagnetic Excitation of Nucleon Resonances,''
\href{https://doi.org/10.1140/epjst/e2011-01488-9}{\B Eur. Phys. J. ST \textbf{198}, 141 (2011).}
%%doi:10.1140/epjst/e2011-01488-9
%%[arXiv:1109.6745 [nucl-th]].
%140 citations counted in INSPIRE as of 24 Oct 2025

 

\bibitem{MAID-website}
  MAID,
  ``Photo- and Electroproduction of Pions, Eta, Etaprime and Kaons on the Nucleon,''
  %\url{https://maid.kph.uni-mainz.de/}
\href{https://maid.kph.uni-mainz.de/}{\B https://maid.kph.uni-mainz.de/.}



\bibitem{Arrington07a}
J.~Arrington, C.~D.~Roberts and J.~M.~Zanotti,
``Nucleon electromagnetic form-factors,''
\href{https://doi.org/10.1088/0954-3899/34/7/S03}{\B J. Phys. G \textbf{34}, S23 (2007).}
%%doi:10.1088/0954-3899/34/7/S03
%%[arXiv:nucl-th/0611050 [nucl-th]].
%235 citations counted in INSPIRE as of 26 Oct 2025

\bibitem{Puckett18a}
A.~J.~R.~Puckett, E.~J.~Brash, M.~K.~Jones, W.~Luo, M.~Meziane, L.~Pentchev, C.~F.~Perdrisat, V.~Punjabi, F.~R.~Wesselmann and A.~Afanasev, \textit{et al.}
``Polarization Transfer Observables in Elastic Electron Proton Scattering at $Q^2 = $2.5, 5.2, 6.8, and 8.5 GeV$^2$,''
\href{https://doi.org/10.1103/PhysRevC.96.055203}{\B 
Phys. Rev. C \textbf{96}, 055203 (2017).}
[erratum: Phys. Rev. C \textbf{98}, 019907 (2018)]
%doi:10.1103/PhysRevC.96.055203
%%[arXiv:1707.08587 [nucl-ex]].
%107 citations counted in INSPIRE as of 26 Oct 2025





\bibitem{N1440-proc}
G.~Ramalho and K.~Tsushima,
``Valence quark contributions for the
$\gamma N \to P_{11}(1440)$ transition,''
\href{https://doi.org/10.1063/1.3647158}{\B 
AIP Conf. Proc. \textbf{1374}, 353 (2011).}
%%doi:10.1063/1.3647158
%%[arXiv:1010.2765 [hep-ph]].
%5 citations counted in INSPIRE as of 22 Oct 2025




\bibitem{Lin08a}
H.~W.~Lin, S.~D.~Cohen, R.~G.~Edwards and D.~G.~Richards,
``First Lattice Study of the $N-P_{11}(1440)$ Transition Form Factors,''
\href{https://doi.org/10.1103/PhysRevD.78.114508}{\B 
Phys. Rev. D \textbf{78}, 114508 (2008).}
%%doi:10.1103/PhysRevD.78.114508
%%[arXiv:0803.3020 [hep-lat]].
%56 citations counted in INSPIRE as of 04 Nov 2025





\bibitem{Gutsche18a}
T.~Gutsche, V.~E.~Lyubovitskij and I.~Schmidt,
``Electromagnetic structure of nucleon and Roper in soft-wall AdS/QCD,''
\href{https://doi.org/10.1103/PhysRevD.97.054011}{\B Phys. Rev. D \textbf{97}, 054011 (2018).}
%%doi:10.1103/PhysRevD.97.054011
%%[arXiv:1712.08410 [hep-ph]].
%37 citations counted in INSPIRE as of 23 Oct 2025





\bibitem{GA-AdS}
G.~Ramalho,
``Holographic estimate of the meson cloud contribution to nucleon axial form factor,''
\href{https://doi.org/10.1103/PhysRevD.97.073002}{\B Phys. Rev. D \textbf{97}, 073002 (2018).}
%%doi:10.1103/PhysRevD.97.073002
%%[arXiv:1707.07206 [hep-ph]].
%10 citations counted in INSPIRE as of 22 Oct 2025




\bibitem{N1535-ST}
G.~Ramalho,
``Improved empirical parametrizations of the $\gamma^\ast N \to N(1535)$ transition amplitudes and the Siegert's theorem,'' 
%Phys. Lett. B \textbf{759}, 126 (2016).
\href{https://doi.org/10.1016/j.physletb.2016.05.060}{\B Phys. Lett. B \textbf{759}, 126 (2016).}
%%doi:10.1016/j.physletb.2016.05.060
%%[arXiv:1602.03444 [hep-ph]].
%19 citations counted in INSPIRE as of 22 Oct 2025


\bibitem{N1520-ST1}
G.~Ramalho,
``Improved empirical parametrizations of the $\gamma^\ast N \to \Delta(1232)$ and $\gamma^\ast N \to N(1520)$ helicity amplitudes and the Siegert's theorem,''
\href{https://doi.org/10.1103/PhysRevD.93.113012}{\B 
Phys. Rev. D \textbf{93}, 113012 (2016).}
%%doi:10.1103/PhysRevD.93.113012
%%[arXiv:1602.03832 [hep-ph]].
%27 citations counted in INSPIRE as of 22 Oct 2025


\bibitem{N1520-ST2}
G.~Ramalho,
``About the magnitude of the  $\gamma^\ast N \to N(1520)$ transverse amplitudes near $Q^2=0$,''
\href{https://doi.org/10.1103/PhysRevD.109.074021}{\B Phys. Rev. D \textbf{109}, 074021 (2024).}
%%doi:10.1103/PhysRevD.109.074021
%%[arXiv:2312.10654 [hep-ph]].
%0 citations counted in INSPIRE as of 22 Oct 2025








\bibitem{HADES2017}
J.~Adamczewski-Musch \textit{et al.} [HADES],
``$\Delta$(1232) Dalitz decay in proton-proton collisions at T=1.25 GeV measured with HADES at GSI,''
\href{https://doi.org/10.1103/PhysRevC.95.065205}{\B Phys. Rev. C \textbf{95}, 065205 (2017).}
%%doi:10.1103/PhysRevC.95.065205
%%[arXiv:1703.07840 [nucl-ex]].
%51 citations counted in INSPIRE as of 23 Oct 2025


\bibitem{HADES2021}
J.~Adamczewski-Musch \textit{et al.} [HADES and PANDA],
``Production and electromagnetic decay of hyperons: a feasibility study with HADES as a phase-0 experiment at FAIR,''
\href{https://doi.org/10.1140/epja/s10050-021-00388-w}{\B 
Eur. Phys. J. A \textbf{57}, 138 (2021).}
%%doi:10.1140/epja/s10050-021-00388-w
%%[arXiv:2010.06961 [nucl-ex]].
%25 citations counted in INSPIRE as of 23 Oct 2025





\bibitem{NSTAR2024}
  G.~Ramalho,
  Contribution to the Workshop NSTAR 2024 (York, UK, 2024). \\
  %\url{https://indico.jlab.org/event/729/}
 \href{https://indico.jlab.org/event/729/}{\B https://indico.jlab.org/event/729/.}





\bibitem{Gegelia16a}
J.~Gegelia, U.~G.~Mei{\ss}ner and D.~L.~Yao,
``The width of the Roper resonance in baryon chiral perturbation theory,''
\href{https://doi.org/10.1016/j.physletb.2016.07.068}{\B Phys. Lett. B \textbf{760}, 736 (2016).}
%%doi:10.1016/j.physletb.2016.07.068
%%[arXiv:1606.04873 [hep-ph]].
%26 citations counted in INSPIRE as of 26 Oct 2025
         %%   baryon chiral perturbation theory







\bibitem{Cutkosky90a}
R.~E.~Cutkosky and S.~Wang,
``Poles of the $\pi N$ $P_{11}$ partial wave amplitude,''
\href{https://doi.org/10.1103/PhysRevD.42.235}{\B Phys. Rev. D \textbf{42}, 235 (1990).}
%%doi:10.1103/PhysRevD.42.235
%82 citations counted in INSPIRE as of 26 Oct 2025
%% Roper  double pole

\bibitem{Arndt06a}
R.~A.~Arndt, W.~J.~Briscoe, I.~I.~Strakovsky and R.~L.~Workman,
``Extended partial-wave analysis of $\pi N$ scattering data,''
\href{https://doi.org/10.1103/PhysRevC.74.045205}{\B Phys. Rev. C \textbf{74}, 045205 (2006).}
%%doi:10.1103/PhysRevC.74.045205
%%[arXiv:nucl-th/0605082 [nucl-th]].
%498 citations counted in INSPIRE as of 26 Oct 2025
%%  GWU   SAID --   Roper  double pole


\bibitem{Arndt85a}
R.~A.~Arndt, J.~M.~Ford and L.~D.~Roper,
%%``PION - NUCLEON PARTIAL WAVE ANALYSIS TO 1100-MeV,''
``Pion-nucleon partial wave analysis to 1100 MeV,''
\href{https://doi.org/10.1103/PhysRevD.32.1085}{\B Phys. Rev. D \textbf{32}, 1085 (1985).}
%%doi:10.1103/PhysRevD.32.1085
%190 citations counted in INSPIRE as of 26 Oct 2025




\bibitem{Sarantsev08a}
A.~V.~Sarantsev, M.~Fuchs, M.~Kotulla, U.~Thoma, J.~Ahrens, J.~R.~M.~Annand, A.~V.~Anisovich, G.~Anton, R.~Bantes and O.~Bartholomy, \textit{et al.}
``New results on the Roper resonance and the $P_{11}$ partial wave,''
\href{https://doi.org/10.1016/j.physletb.2007.11.055}{\B 
Phys. Lett. B \textbf{659}, 94 (2008).}
%%doi:10.1016/j.physletb.2007.11.055
%%[arXiv:0707.3591 [hep-ph]].
%125 citations counted in INSPIRE as of 26 Oct 2025



\bibitem{DRoberts14a}
D.~S.~Roberts, W.~Kamleh and D.~B.~Leinweber,
``Nucleon Excited State Wave Functions from Lattice QCD,''
\href{https://doi.org/10.1103/PhysRevD.89.074501}{\B 
Phys. Rev. D \textbf{89}, 074501 (2014).}
%%doi:10.1103/PhysRevD.89.074501
%%[arXiv:1311.6626 [hep-lat]].
%37 citations counted in INSPIRE as of 27 Oct 2025
    %%   Roper WF in Lattice




\bibitem{Mathur05a}
N.~Mathur, Y.~Chen, S.~J.~Dong, T.~Draper, I.~Horvath, F.~X.~Lee, K.~F.~Liu and J.~B.~Zhang,
``Roper resonance and $S_{11}$(1535) from lattice QCD,''
\href{https://doi.org/10.1016/j.physletb.2004.11.010}{\B Phys. Lett. B \textbf{605}, 137 (2005).}
%%doi:10.1016/j.physletb.2004.11.010
%%[arXiv:hep-ph/0306199 [hep-ph]].
%174 citations counted in INSPIRE as of 17 Nov 2025





\bibitem{Mahbub10b}
M.~S.~Mahbub, W.~Kamleh, D.~B.~Leinweber, A.~O Cais and A.~G.~Williams,
``Ordering of Spin-$\frac{1}{2}$ Excitations of the Nucleon in Lattice QCD,''
\href{https://doi.org/10.1016/j.physletb.2010.08.049}{\B Phys. Lett. B \textbf{693}, 351 (2010).}
%%doi:10.1016/j.physletb.2010.08.049
%%[arXiv:1007.4871 [hep-lat]].
%37 citations counted in INSPIRE as of 17 Nov 2025




\bibitem{Lang17a}
C.~B.~Lang, L.~Leskovec, M.~Padmanath and S.~Prelovsek,
``Pion-nucleon scattering in the Roper channel from lattice QCD,'' \\
\href{https://doi.org/10.1103/PhysRevD.95.014510}{\B 
Phys. Rev. D \textbf{95}, 014510 (2017).}
%%doi:10.1103/PhysRevD.95.014510
%[arXiv:1610.01422 [hep-lat]].
%117 citations counted in INSPIRE as of 18 Nov 2025


\bibitem{Sun20a}
M.~Sun \textit{et al.} [xQCD],
``Roper State from Overlap Fermions,'' \\
\href{https://doi.org/10.1103/PhysRevD.101.054511}{\B Phys. Rev. D \textbf{101}, 054511 (2020).}
%%doi:10.1103/PhysRevD.101.054511
%%[arXiv:1911.02635 [hep-ph]].
%18 citations counted in INSPIRE as of 17 Nov 2025




\bibitem{Wu18b}
J.~j.~Wu, D.~B.~Leinweber, Z.~w.~Liu and A.~W.~Thomas,
``Structure of the Roper Resonance from Lattice QCD Constraints,'' \\
\href{https://doi.org/10.1103/PhysRevD.97.094509}{\B Phys. Rev. D \textbf{97}, 094509 (2018).}
%%doi:10.1103/PhysRevD.97.094509
%%[arXiv:1703.10715 [nucl-th]].
%58 citations counted in INSPIRE as of 24 Oct 2025


\bibitem{Owa25a}
S.~Owa, D.~B.~Leinweber and A.~W.~Thomas,
``Nucleon resonance structure up to 2~GeV and the nature of the Roper resonance,''
\href{https://doi.org/10.1103/tt7s-p9gj}{\B Phys. Rev. D \textbf{111}, 116002 (2025).}
%%doi:10.1103/tt7s-p9gj
%%[arXiv:2503.09945 [hep-ph]].
%4 citations counted in INSPIRE as of 24 Oct 2025



\bibitem{Liu16a}
Z.~W.~Liu, W.~Kamleh, D.~B.~Leinweber, F.~M.~Stokes, A.~W.~Thomas and J.~J.~Wu,
``Hamiltonian effective field theory study of the $\mathbf{N^*(1535)}$ resonance in lattice QCD,''
\href{https://doi.org/10.1103/PhysRevLett.116.082004}{\B Phys. Rev. Lett. \textbf{116}, 082004 (2016).}
%%doi:10.1103/PhysRevLett.116.082004
%%[arXiv:1512.00140 [hep-lat]].
%83 citations counted in INSPIRE as of 24 Oct 2025

  

\bibitem{Liu17a}
Z.~W.~Liu, W.~Kamleh, D.~B.~Leinweber, F.~M.~Stokes, A.~W.~Thomas and J.~J.~Wu,
``Hamiltonian effective field theory study of the $\mathbf{N^*(1440)}$ resonance in lattice QCD,''
\href{https://doi.org/10.1103/PhysRevD.95.034034}{\B Phys. Rev. D \textbf{95}, 034034 (2017).}
%%doi:10.1103/PhysRevD.95.034034
%%[arXiv:1607.04536 [nucl-th]].
%78 citations counted in INSPIRE as of 18 Nov 2025




\bibitem{Workman12a}
R.~L.~Workman, R.~A.~Arndt, W.~J.~Briscoe, M.~W.~Paris and I.~I.~Strakovsky,
``Parameterization dependence of $T$ matrix poles and eigenphases from a fit to $\pi$N elastic scattering data,''
\href{https://doi.org/10.1103/PhysRevC.86.035202}{\B Phys. Rev. C \textbf{86}, 035202 (2012).}
%%doi:10.1103/PhysRevC.86.035202
%%[arXiv:1204.2277 [hep-ph]].
%142 citations counted in INSPIRE as of 18 Nov 2025


\bibitem{GWU}
   %R.~A.~Arndt, W.~J.~Briscoe, I.~I.~Strakovsky, R.~L.~Workman,
  ``JBW Interactive Scattering Analysis website,''
  \href{https://JBW.phys.gwu.edu}{\B https://JBW.phys.gwu.edu.}



\end{thebibliography}
\end{document}